\documentclass[aps,prd,twocolumn,showpacs,superscriptaddress,groupedaddress]{revtex4-2}  
\usepackage[T1]{fontenc}
\usepackage{lmodern} 
\usepackage{graphicx}  
\usepackage{dcolumn}   
\usepackage{bm}        
\usepackage{amssymb}   
\usepackage{amsmath}
\usepackage{bbold}
\usepackage{array}
\usepackage{makecell}
\usepackage{braket}
\usepackage{longtable}
\usepackage{supertabular,booktabs}

\usepackage{titlesec} 
\usepackage[colorlinks,citecolor=blue,urlcolor=blue,hypertexnames=true]{hyperref}
\usepackage{subfigure}

\usepackage[usenames,dvipsnames,svgnames,table]{xcolor} 

\newcommand{\be}{\begin{equation}}
\newcommand{\ee}{\end{equation}}
\newcommand{\bea}{\begin{eqnarray}}
\newcommand{\eea}{\end{eqnarray}}
\newcommand{\bml}{\begin{subequations}}
\newcommand{\eml}{\end{subequations}}
\newcommand{\bfig}{\begin{figure}}
\newcommand{\efig}{\end{figure}}

\newcommand{\del}{\delta}

\newcommand{\bmat}{\begin{pmatrix}}
\newcommand{\emat}{\end{pmatrix}}

\begin{document}

\widetext


\title{\textcolor{Sepia}{\textbf \huge Entanglement in interacting quenched two-body coupled oscillator system}}

\author{{\large  Sayantan Choudhury${}^{1}$}}
\email{sayantan.choudhury@icts.res.in\\  sayanphysicsisi@gmail.com}

\author{ \large Rakshit Mandish Gharat${}^{2}$}
\email{rakshitmandishgharat.196ph018@nitk.edu.in }
\author{\large Saptarshi Mandal${}^{3}$}
\email{saptarshijhikra@gmail.com }
\author{ \large Nilesh Pandey${}^{4}$}
\email{nilesh911999@gmail.com  }
\author{ \large Abhishek Roy${}^{5}$}
\email{roy.1@iitj.ac.in }
\author{\large Partha Sarker${}^{6}$\vspace{.4cm}}
\email{sarker239@gmail.com }

\affiliation{ ${}^{1}$International Centre for Theoretical Sciences, Tata Institute of Fundamental Research (ICTS-TIFR), Shivakote, Bengaluru 560089, India,}
\affiliation{${}^{2}$Department of Physics, National Institute of Technology Karnataka, Surathkal, Karnataka-575025, India,}
\affiliation{${}^{3}$ Department of Physics, Indian Institute of Technology Kharagpur, Kharagpur-721302, India,}
\affiliation{${}^{4}$Department of Applied Physics, Delhi Technological University, Delhi-110042, India,}
\affiliation{${}^{5}$ Department of Physics, Indian Institute of Technology Jodhpur,Karwar, Jodhpur - 342037, India.}
\affiliation{${}^{6}$Department of Physics, University of Dhaka, Curzon Hall, Dhaka 1000, Bangladesh.}

\begin{abstract}
{In this work, we explore the effects of a quantum quench on the entanglement measures of a two-body coupled oscillator system having quartic interaction. We use the invariant operator method, under a perturbative framework, for computing the ground state of this system}. We give the analytical expressions for the total and reduced density matrix of the system having non-Gaussian, quartic interaction terms. Using this reduced density matrix, we show the analytical calculation of two entanglement measures viz., Von Neumann entanglement entropy using replica trick and Renyi entanglement entropy. Further, we give a numerical estimate of these entanglement measures with respect to the dimensionless parameter $(t/\delta t$) and show its behaviour in the three regimes, i.e; late time behaviour, around the quench point and the early time behaviour.
We comment on the variation of these entanglement measures for different orders of coupling strength. The variation of Renyi entropy of different orders has also been discussed. 
\end{abstract}

\pacs{}
\maketitle
\section{\textcolor{Sepia}{\textbf{ \large  Introduction}}}
\label{sec:introduction}


In recent years, the most important works in theoretical physics have been studied by blending the ideas of quantum many-body physics, quantum information theory and quantum field theory. The amalgamation of these branches of physics have been reviewed in \cite{intro:merge1,intro:merge3,intro:merge4,intro:merge5,intro:merge6,intro:merge7,intro:merge8}. These works mostly focus on studying the dynamical properties of quantum entanglement in time-dependent systems \cite{intro:time-ent1,intro:time-ent2,intro:time-ent3,intro:time-ent4,intro:time-ent5,intro:time-ent7,intro:time-ent8,ent1,ent2,ent3}. This dynamical entanglement can be characterised by computing von Neumann entanglement entropy and Renyi entanglement entropy. The traditional way of computing these entanglement measures involves constructing the reduced density matrix using the eigenstates of the time-dependent Hamiltonian.

One of the ways to compute the eigenstates of such time-dependent Hamiltonians is by constructing the Lewis-Resenfield invariant operator and this is often termed as invariant operator representation of the wavefunction \cite{doi:10.1063/1.1664991}. Some works following this approach to compute the time-dependent eigenstates are \cite{intro:invar1,intro:invar2,invarSC,intro:invar3,intro:invar4}. Evolution of such time-dependent quantum states can be studied as solution to the Ermakov-Milne-Pinney equation \cite{ermakov2008second,milne1930numerical,pinney1950nonlinear}. The invariant operator method is generalised for perturbed theories by computing time-independent perturbative corrections \cite{2020FrP.....8..189C}, considering adiabatic evolution \cite{intro:pert-adiab1,intro:pert-adiab2} of the parameters.

For unperturbed Hamiltonians or free theories, the reduced density matrix once computed represents the Gaussian states. Entanglement can then be computed using these Gaussian states \cite{intro:Gaussianent2,intro:Gaussianent3,Gaussianent4,Gaussianent4}. Computing von Neumann entropy for these Gaussian states employs the use of correlation functions defined using the framework of quantum field theory \cite{intro:Gaussianent1}.

Entanglement in interacting theories has been studied using replica trick \cite{1994NuPhB.424..443H} in \cite{2020JHEP...11..114C}.  On the other hand, the perturbed entanglement entropy is computed using the path-integral approach in \cite{perturbedent1}.

Most of the recent works in many-body physics have been about contemplating the behaviour of entanglement for a system having a time-dependent parameter in Hamiltonian. This time-dependent parameter can be varied quickly or slowly and hence the process is termed as a "quantum quench". Some of the most important works for studying entanglement-properties of extended systems undergoing a quantum quench are \cite{intro:quench1,intro:quench2,intro:quench3,intro:quench4,intro:quench5,intro:quench6,intro:quench7,intro:quench8,Ghosh:2017nlk,Ghosh:2019yjh}. These quantum quenches can be thought of as protocols driving the system out-of-equilibrium \cite{intro:ooe1,intro:ooe2,intro:ooe3}. These local equilibrium can then be analysed in quenched systems by using reduced density matrix. In recent years, effects of quantum quenches have even been studied experimentally using cold atom systems\cite{e1,e2,e3,e4,e5,e6,e7,e8,e9,e10}. Studying the effect of quantum quench in the case of interacting theories or perturbed theories is of prime importance \cite{intro:quenchInt1}.

{In recent years, the study of coupled oscillators has been an area of active research. This is primarily due to the extensive use of such models in studying quantum and non-linear physics \cite{Coup11,Coup12,Coup13,Coup14,Coup15}, molecular chemistry\cite{Coup21,COup22,Coup23} and biophysics\cite{Coup31,COup32,Coup33}. Especially, in quantum physics, analysing entanglement of coupled oscillators is of prime importance.\cite{Coupe1,Coupe2,Coupe3,Coupe4,Coupe5}
}

{Motivated by the discussion given above, in this work we consider a toy model with a Hamiltonian of two coupled oscillators having quartic self-coupling term. The Hamiltonian for this system consists of a time-dependent quench profile.  The eigenstates for this time-dependent Hamiltonian are computed using invariant operator method, in a perturbative framework}. Further, the reduced density matrix (with quartic interaction terms) is constructed using time-independent perturbation theory. The dynamical von Neumann entropy and Renyi entropy are then derived using appropriate formulae for the obtained reduced density matrix. We comment on the behaviour of these entanglement measures by varying the relevant parameters.

The organisation of the paper is as follows:
\begin{itemize}
    \item We start our discussion by providing an overview of the quench protocol and Hamiltonian of the system in section \ref{sec:QS}. 
    \item In section \ref{sec:E2}, the expression for eigenstates of the time-dependent Hamiltonian is computed using invariant-operator representation of wavefunction. Further, the expression for first order time-independent correction to the ground state of the Hamiltonian is also approximated in this section.
    \item In section \ref{v2b}, we use the ground state wavefunction with perturbative correction to compute the expression for reduced density matrix, with quartic interaction terms. von Neumann entanglement entropy is then computed by performing the replica trick over this reduced density matrix. Further, we show the analytically computed expression for Renyi entanglement entropy.
    \item In section \ref{sec:numerical}, we numerically evaluate the respective entanglement measures and plot them with respect to the dimensionless parameter $(t/\delta t)$. We comment on the parametric variation of these entanglement measures for each of the chosen three regimes.
    \item Section \ref{sec:Conclusions} summarises the conclusions we draw from the obtained results of this work with some interesting future prospects of our present work. 
\end{itemize}

\section{\textcolor{Sepia}{\textbf{\large The Setup and the Quench protocol}}}\label{sec:QS} 
{In this section we begin by discretising the Hamiltonian for a scalar field theory with $\phi^4$ interaction term on a lattice. We show that the Hamiltonian then represents a family of infinite anharmonic oscillators with quartic couplings. In this article, we study a system of two coupled oscillators having quartic perturbation}. Furthermore, we use normal mode basis to decouple the Hamiltonian so that we can compute the eigenstates for this system in a much simpler way, in upcoming section. Also, we mention the time-dependent quench profile chosen as the frequency of this Hamiltonian.\\

The Hamiltonian for a scalar field theory with a $\hat{\lambda}\phi^4$ interaction is given by \cite{fieldH}, 
 \begin{align}
 \mathcal{H}=\frac{1}{2} \int d^{d-1} x&\Bigg[\pi(x)^{2}+(\nabla \phi(x))^{2}+m^{2} \phi(x)^{2}\nonumber\\
 &+\frac{\hat{\lambda}}{12}\phi(x)^{4}\Bigg]~. 
\end{align}
Here $d$ is the space-time dimensions. We assume that the coupling $\hat{\lambda}<<1$, so that we can work in a perturbative framework. This theory can be discretized on a $d-1$ dimensional lattice, {which is characterised by lattice spacing, $\delta$.} 
It can be shown that, the discretized Hamiltonian becomes,
\begin{align}
\mathcal{H}=\frac{1}{2} &\sum_{\vec{n}}\Bigg\{\frac{\pi(\vec{n})^{2}}{\delta^{d-1}}+\delta^{d-1}\Bigg[\frac{1}{\delta^{2}} \sum_{i}\left(\phi(\vec{n})-\phi\left(\vec{n}-\hat{x}_{i}\right)\right)^{2}\nonumber\\
&+m^{2} \phi(\vec{n})^{2}+\frac{\hat{\lambda}}{12}\phi(\vec{n})^{4}\Bigg]\Bigg\}.
\end{align}
Here $\vec{n}$ denotes the spatial location of the points on lattice and $\hat{x}_{i}$ represent the unit vectors along the lattice. Further, we introduce the following substitutions to simplify the form of the Hamiltonian:
\begin{align}
\hat{X}(\vec{n})&=\delta^{d / 2} \phi(\vec{n}), & \hat{P}(\vec{n})&=\pi(\vec{n}) / \delta^{d / 2}, &\nonumber\\ M&=\frac{1}{\delta}, &\omega=m ,& \nonumber\\ 
\eta&=\frac{1}{\delta}, &\lambda=\frac{\hat{\lambda}}{24} \delta^{-d},
\end{align}
{where $\omega$ represents the frequency of individual oscillators and $\eta$ denotes inter-mass coupling.} After these substitutions we get,
\begin{equation}\label{Eq_3.3}
\begin{split}
\mathcal{H}={}&\sum_{\vec{n}}\Big\{\frac{\hat{P}(\vec{n})^{2}}{2 M}+\frac{1}{2} M\Big[\omega^{2} \hat{X}(\vec{n})^{2}\\&+\eta^{2} \sum_{i}\left(\hat{X}(\vec{n})-\hat{X}\left(\vec{n}-\hat{x}_{i}\right)\right)^{2}\\
&+2\lambda \hat{X}(\vec{n})^{4}\Big]\Big\}.
\end{split}
\end{equation}
The above Hamiltonian, in Eq. \eqref{Eq_3.3} represents a family of infinite coupled anharmonic  oscillators. In this work we focus on the system representing two coupled oscillators and compute the entanglement for this system. {Setting $M=1$, for simplicity, Eq. \eqref{Eq_3.3} can be specialised for case of two coupled oscillators},
\begin{equation}\label{Eq_3.4}
\begin{aligned}
H= \frac{1}{2}\Big[p_{1}^{2}+p_{2}^{2}+\omega^{2}\left(x_{1}^{2}+x_{2}^{2}\right)+\eta^{2}\left(x_{1}-x_{2}\right)^{2}\\+2\left\{ \lambda\left(x_{1}^{4}+x_{2}^{4}\right)\right\}\Big].
\end{aligned}
\end{equation}
Here $x_i$ and $p_i$, for $i=1,2$ denote the cannonical coordinates of the respective oscillator following the standard commutation relation $[x_i,p_j]=i\delta^i_j$, while $\lambda$ denotes the coupling coefficient of $\phi^4$ interaction term.\\
The eigenstates of the above Hamiltonian Eq. \eqref{Eq_3.4}, can easily be computed using normal coordinates defined as,
\begin{equation}\label{coord}
\begin{aligned}
X_1=({x_1}+{x_2})/\sqrt{2} \\
X_2=({x_1}-{x_2})/\sqrt{2} \\
P_1=({p_1}+{p_2})/\sqrt{2} \\ 
P_2=({p_1}-{p_2})/\sqrt{2} .
\end{aligned}
\end{equation}
The unperturbed part of Hamiltonian when written using these normal coordinates decouples. One can then show that the total Hamiltonian of Eq. \eqref{Eq_3.4} in normal coordinates takes the following form:
\bea
H=\sum_{i=1}^2H_i+H_p\hspace{5pt},\nonumber
\eea
where,
\bea
\label{decoup}
H_i(T)=\frac{1}{2}\biggr({P_i}^2+\omega_i^2(T){X_i}^2\biggr),
\eea
denotes the unperturbed and decoupled Hamiltonian for each of the two oscillators. On the other hand the perturbed Hamiltonian is given by,
\begin{equation}\label{Hp}
\begin{aligned}
    H_p&=\lambda V=\lambda(x_{1}^{4}+x_{2}^{4})\\
   &=\lambda(X_{1}^{4}+X_{2}^{4}+6X_{1}^{2}X_{2}^{2}).
\end{aligned}
\end{equation}
This enables us to use $\lambda\phi^4$ time-indepedent perturbation theory and compute the eigenstates of total Hamiltonian of Eq. \eqref{Eq_3.4}.\\
We now consider the frequency $\omega$ in, Eq. \eqref{Eq_3.4} as a time-dependent quench profile.
One of the most common quench profiles used in literature \cite{Caputa:2017ixa,PhysRevLett.122.081601} is given by:
\be\label{quenchprof}
\omega^2(t/\delta t)=\omega_0^2\left[\tanh^2{\left(\frac{t}{\delta t}\right)}\right].
\ee  
Here $\omega_0$ can be interpreted as a free parameter and $\delta t$ is the quench parameter or the quench rate. The quench profile chosen here is such that it admits an exact solution for the mode functions given in \cite{PhysRevLett.122.081601} and the quench profile attains a constant value at late and early times. The dynamical process due to this profile happens in the $[-\delta t,\delta t]$ time window. We will set $t/\delta t=T$ and $\omega_0=1$. 
The respective frequencies in normal mode basis take the following form,
\bea
\omega_1=\omega(T)~\text{and}~\omega_2=\sqrt{\omega^2(T)+4\eta^2}\hspace{3pt}.
\eea
where $\omega(T)$ is the quench profile Eq. (\ref{quenchprof}).\\
Note that the unperturbed Hamiltonian of Eq. \eqref{decoup} is now time-dependent while the perturbed Hamiltonian of Eq. \eqref{Hp} acts as time-independent $\phi^4$ coupling applied on the two harmonic oscillators. In section \ref{sec:E2}, we show the analytical computation of ground state, $\Psi_{0,0}$ of the total Hamiltonian of two coupled anharmonic oscillators having $\lambda\phi^4$ perturbation. This ground state is used to derive the analytical expressions of the respective entanglement measures in section \ref{v2b}.
\section{\textcolor{Sepia}{\textbf{\large Constructing Wave function for a $\phi^4$ quench model}}} 
\label{sec:E2}
In this section our prime objective is to construct the wavefunction approximated to first order in coupling constant $\lambda$. In subsection \ref{unper} we compute the eigenstates of decoupled and unperturbed Hamiltonian Eq. \eqref{decoup}. These eigenstates are then used to construct the ground state of perturbed Hamiltonian Eq. \eqref{Hp}, approximated to first perturbative order, in subsection \ref{perturbed}. Finally we compute the  total wavefunction as ground state of total Hamiltonian Eq. \eqref{Eq_3.4}.

\subsection{Eigenstates and Eigenvalues for unperturbed Hamiltonian}\label{unper}
As, the unperturbed Hamiltonian Eq. \eqref{decoup} decouples in the normal mode basis, the eigenstates for the unperturbed Hamiltonian are just the product of the eigenstates of respective oscillators in the normal-mode basis: \be
\psi^{(0)}_{n_1,n_2}(X_1,X_2,T)=\psi_{n_1}(X_1,T)\psi_{n_2}(X_2,T).
\ee
Since the unperturbed Hamiltonian consists of a time-dependent frequency scale, we now use a prescription often termed as the invariant representation in the literature \cite{1994PhRvA..50.1035Y}, to get the unperturbed eigenstates.  
  \\
We begin the invariant representation by listing the auxiliary equations. The solutions to these equations can then be used to compute the coupled wavefunction. The auxiliary equations can be written as:
\begin{equation}\label{aux}
\begin{aligned}
\ddot{\sigma}_j-\sigma_j\dot{\gamma_j}+\omega_j^2(T)\sigma_j=0\\
\sigma_j\ddot{\gamma}_j+2\dot{\sigma}_j\dot{\gamma}_j=0.
\end{aligned}
\end{equation}
Here,  j=1,2 and $\sigma_j(T)$ and $\gamma_j(T)$ are time-dependent factors for each of the two coupled oscillators. Also, $\dot{\gamma}_j=\partial_T\gamma_j$, $\dot{\sigma}_j=\partial_T\sigma_j$ and $\ddot{\sigma}_j=\partial^2_T\sigma_j$ . The subscript $j$ denotes the oscillator for which the respective parameter is mentioned. The computation of explicit values of $\sigma(T)$ and $\gamma(T)$ is discussed in appendix \ref{MPequation}. Note that we have suppressed the time-dependence throughout this section. \\
Next, we define the creation $(a^\dagger_j)$ and annihilation $(a_j)$ operators given by,
\begin{equation}\label{crea}
\begin{aligned}
 a_j=\frac{1}{\sqrt{2\dot{\gamma_j}}}\biggr[\dot{\gamma}_j\biggr(1-i\frac{\dot{\sigma_j}}{\sigma_j\dot{\gamma_j}}\biggr)X_j+iP_j\biggr] \\
 a_j^\dagger=\frac{1}{\sqrt{2\dot{\gamma_j}}}\biggr[\dot{\gamma}_j\biggr(1+i\frac{\dot{\sigma_j}}{\sigma_j\dot{\gamma_j}}\biggr)X_j-iP_j\biggr].\\
\end{aligned}
\end{equation}
Here, j=1,2. One can show that these operators satisfy the commutation relation $[a_i,a_j^\dagger]=\delta_j^i$.
The creation and annihilation operators can be used to define invariant operator for the respective decoupled Hamiltonian,
\bea\label{invar}
I_j=\Omega_j\biggr(a_j^\dagger a_j+\frac{1}{2}\biggr).
\eea
Here, j=1,2. On the other hand, $\Omega_j=\sigma_j^2\dot{\gamma_j}$, is an invariant quantity with respect to time. The construction of this invariant operator Eq. \eqref{invar} has been briefly outlined in appendix \ref{appendixA}. The invariant operator has its own spectrum and eigenstates. The eigenstates of invariant operator can be used to formulate the wavefunctions for each decoupled Hamiltonian. The outline of the same is given in appendix \ref{appendixA} . Using equation (\ref{eigen}) for ${n_1}, {n_2}=0$, one can show that the ground state of unperturbed Hamiltonian is given by,
\begin{equation}\label{gnd}
 \begin{aligned}
   {\psi_{0,0}^{(0)}}&=\sqrt{\frac{g_{1} g_{2}}{\pi}}\exp\left[-i\frac{\gamma_1+\gamma_2}{2}\right]\times\\
  & \exp\left[-\frac{1}{2}\Big(g_1^2(1-id)X_1^2+g_2^2(1-if)X_2^2\Big)\right].
 \end{aligned}
\end{equation}

where, the coefficients $g_1$, $g_2$, $d$ and $f$ are given by,
\bea\label{d}
g_1=\sqrt{\dot{\gamma_1}}, ~g_2=\sqrt{\dot{\gamma_2}},~d=\frac{\dot{\sigma_1}}{\dot{\gamma_1}\sigma_1},~f=\frac{\dot{\sigma_2}}{\dot{\gamma_2}\sigma_2}.
\eea
Next we emphasize that the eigenvalues of the unperturbed decoupled Hamiltonians in Eq. \eqref{decoup} will have a time dependent factor \cite{2020FrP.....8..189C}. These eigenvalues for each of the decoupled Hamiltonians are given as:
\begin{equation}
 \begin{aligned}
    \bra{\psi_{n_j}}H_i\ket{\psi_{n_j}}&=W_j(T)\left[n_j+\frac{1}{2}\right],
 \end{aligned}
\end{equation}
where, $j=1,2$. Here $W_j(T)$ is the time-dependent factor for each oscillator given by,
\be
W_j(T)=\frac{\dot{\gamma_j}}{2}\left(\frac{\dot{\sigma_i}+\sigma_i^2\omega_i^2+\sigma_i^2\dot{\gamma_i}}{\sigma_i^2\dot{\gamma_i}^2}\right),
\ee
where $j=1,2$. Using the above eigenvalues one can write the energy eigenvalue for the unperturbed state of two coupled oscillators, Eq. (\ref{eigen}) as:
\bea
\bra{\psi_{n_1,n_2}^{(0)}}H\ket{\psi_{n_1,n_2}^{(0)}}&&=W_1(T)\left(n_1+\frac{1}{2}\right)\nonumber\\
&&+W_2(T)\left(n_2+\frac{1}{2}\right).
\label{evalue}
\eea

\subsection{Ground state of two coupled oscillators with first order-$\phi^4$ perturbation}\label{perturbed}
Using time-independent perturbation theory, one can show that the first order perturbative correction to the ground state of two-coupled oscillators is,
\be\label{PC}
\psi_{0,0}^{(1)}=\sum_{(n_1,n_2)\neq(0,0)}\frac{\bra{\psi_{n_1,n_2}^{(0)}}V\ket{\psi_{0,0}^{(0)}}\times\psi_{n_1,n_2}^{(0)}.}{\bra{\psi_{0,0}^{(0)}}H\ket{\psi_{0,0}^{(0)}}-\bra{\psi_{n_1,n_2}^{(0)}}H\ket{\psi_{n_1,n_2}^{(0)}}}
\ee
Using the form of perturbed Hamiltonian Eq. \eqref{Hp} and the time dependent eigenvalues Eq. \eqref{evalue}, the above expression when evaluated becomes,
\begin{align}\label{pc}
  \psi_{0,0}^{(1)}=&-\frac{3(g_1^2+g_2^2)\psi_{0,2}^{(0)}}{4\sqrt{2}hg_1^2g_2^2}-\frac{3 \psi_{0,4}^{(0)}}{8\sqrt{2}hg_2^2}-\frac{3(g_1^2+g_2^2)\psi_{2,0}^{(0)}}{4\sqrt{2}gg_1^2g_2^2}\nonumber\\
 &-\frac{3\psi_{2,2}^{(0)}}{2(2g+2h)2g_1^2g_2^2}-\frac{3\psi_{4,0}^{(0)}}{8\sqrt{2}gg_1^2}.
\end{align}
Here,
\be
\begin{aligned}
g=\bigg(\frac{\dot{\sigma_1}+\sigma_1^2\omega_1^2+\Omega_1\dot{\gamma_1}}{\Omega_1}\bigg),\\
h=\bigg(\frac{\dot{\sigma_2}+\sigma_2^2\omega_2^2+\Omega_2\dot{\gamma_2}}{\Omega_2}\bigg).
\end{aligned}
\ee
The explicit form of the first order correction can be computed using the expression of unperturbed eigenstates, Eq. \eqref{eigen}.
The total wavefunction for ground state of total Hamiltonian Eq. \eqref{Eq_3.4}, corrected to first order of time-independent $\lambda\phi^4$ perturbation, is given by: $\Psi_{0,0}=\psi_{0,0}^{(0)}+\lambda\psi_{0,0}^{(1)}$. Using Eq. \eqref{gnd} and Eq. \eqref{pc} while approximating, the coupling constant $\lambda<<1$ we can express the final form of the wavefunction in normal mode basis as:
\bea\label{wavef}
\nonumber  \Psi_{0,0}(X_1,X_2)&&=\biggr(\frac{g_1^2g_2^2}{\pi^2}\biggr)^\frac{1}{4}e^{-i\frac{(\gamma_1+\gamma_2)}{2}}\exp\biggr[-\frac{1}{2}(1-id)g_1^2X_1^2\\ \nonumber
  &&-\frac{1}{2}(1-if)g_2^2X_2^2+\lambda\left(A_1+A_2X_1^2+A_3X_2^2\right.\\&&\left.+A_4X_1^4+A_5X_2^4+A_6X_1^2X_2^2\right)\biggr].
\eea

The coefficients $A_i$ for $i=1$ to $i=6$ are mentioned in a table given in appendix \ref{app:Table}.
The above wavefunction Eq. \eqref{wavef} represents the ground state of total Hamiltonian in Eq. \eqref{Eq_3.4}, of the system of two coupled oscillators with $\lambda\phi^4$ perturbation.
We take note of the fact that all variables, aside from the coordinates, $X_1,X_2$ and coupling constant, $\lambda$ in the wavefunction, Eq. \eqref{wavef} are functions of timescale $T$. The wavefunction is then dependent on both $t$ and $\delta t$. This explicit time dependence can be evaluated by computing $\sigma_i$ and $\gamma_i$, shown in appendix \ref{MPequation} .


\section{\textcolor{Sepia}{\textbf{\large Analytical calculation of Entanglement Measures}}}\label{v2b}
In the previous sections, \ref{sec:QS} and \ref{sec:E2}, we computed the ground state wavefunction for a system of two coupled bosonic oscillators with a $\phi^4$ first-order perturbative correction for Hamiltonian having a quenched frequency-profile. In this section our prime objective is to show the analytical steps to calculate entanglement measures, viz., von Neumann entanglement entropy and Renyi entropy.\\
In subsection \ref{sub:RDM} reduced density matrix for the system of two coupled oscillators is constructed using the wavefunction Eq. \eqref{wavef}. To compute von Neumann entropy using replica trick \cite{1994NuPhB.424..443H}, \cite{2020JHEP...11..114C} as well as Renyi entropy, the trace of reduced density matrix should be evaluated, this is shown in subsection \ref{trace}. Finally using the appropriate formulae we show the computation of the respective entanglement measures in \ref{sec:EntM}.

\subsection{Density Matrix for Perturbed wavefunction}\label{sub:RDM}
We begin by transforming the wavefunction $\Psi(X_1,X_2)$ given in Eq. \eqref{wavef} to $\Psi(x_1,x_2)$ i.e. we transform the normal coordinates back to space-time coordinates using Eq. \eqref{coord}. We mention four new symbols:
\bea\label{coeff1}
P=\frac{1}{2}(1-id)g_1^2~~;~~P^*=\frac{1}{2}(1+id)g_1^2\nonumber\\
Q=\frac{1}{2}(1-if)g_2^2~~;~~ Q^*=\frac{1}{2}(1+if)g_1^2.
\eea
The wavefunction in spacetime coordinates is then represented by:
\begin{widetext}
	\be\label{e1}  
	\begin{matrix}
\displaystyle   {\Psi}(x_1,x_2)=\left(\frac{g_1^2g_2^2}{\pi^2}\right)^\frac{1}{4}e^{-i(\gamma_1+\gamma_2)/2}\exp\biggr\{-\frac{P}{2}(x_1^2+x_2^2+2x_1x_2)-\frac{Q}{2}(x_1^2+x_2^2-2x_1x_2)+\lambda\biggr[A_1+\frac{A_2}{2}(x_1^2+x_2^2+2x_1x_2) \\ 
 \displaystyle   +\frac{A_3}{2}(x_1^2+x_2^2-2x_1x_2)+\frac{A_4}{4}(x_1^4+x_2^4+4x_1^3x_2+4x_1x_2^3+6x_1^2x_2^2)\\
 \displaystyle   +\frac{A_5}{4}(x_1^4+x_2^4-4x_1^3x_2-4x_1x_2^3+6x_1^2x_2^2)+\frac{A_6}{4}(x_1^4+x_2^4-2x_1^2x_2^2)\biggr]\biggr\}. 
\end{matrix}\ee
	\end{widetext}
The complex conjugate of the above given wavefunction is denoted by $\Psi^*(x_1',x_2')$. Using the conjugate of the wavefunction in Eq. \eqref{e1}, we can construct the total density matrix for the system of two oscillators by ${\rho}(x_1,x_2,x_1',x_2')=\Psi(x_1,x_2)\Psi^*(x_1',x_2')$. One can easily show that the density matrix is given as:
\begin{widetext}
	\be\label{RDM}  
	\begin{matrix}
\displaystyle 	{\rho}(x_1,x_2,x_1',x_2')=\displaystyle\biggr(\frac{g_1g_2}{\pi}\biggr)\exp\biggr\{-\frac{P}{2}(x_1^2+x_2^2+2x_1x_2)-\frac{P^*}{2}(x_1'^2+x_2'^2+2x_1'x_2')\\ 
\displaystyle\quad \quad\quad \quad-\frac{Q}{2}(x_1^2+x_2^2-2x_1x_2)-\frac{Q^*}{2}(x_1'^2+x_2'^2-2x_1'x_2')\\ 
\displaystyle 	+\lambda\biggr[2A_1+\frac{A_2}{2}(x_1^2+x_1'^2+x_2^2+x_2'^2+2x_1 x_2+2x_1' x_2') + \frac{A_3}{2}(x_1^2+x_1'^2+x_2^2+x_2'^2-2x_1 x_2-2x_1' x_2')\\ +\frac{A_4}{4}(x_1^4+x_1'^4+x_2^4+x_2'^4+4x_1^3 x_2+4x_1'^3 x_2'+4x_1 x_2^3+4x_1' x_2'^3+6x_1^2 x_2^2+6x_1'^2 x_2'^{2})\\ \displaystyle 
	+\frac{A_5}{4}(x_1^4+x_1'^4+x_2^4+x_2'^4-4x_1^3 x_2-4x_1'^3 x_2'-4x_1 x_2^3-4x_1' x_2'^3+6x_1^2 x_2^2+6x_1'^2 x_2'^{2})\\ \displaystyle 
   +\frac{A_6}{4}(x_1^4+x_1'^4+x_2^4+x_2'^4-2x_1^2x_2^2-2x_1'^2x_2'^2)\biggr]\biggr\}.
\end{matrix}~\ee
	\end{widetext}
The reduced density matrix can be computed using total density matrix, shown in Eq. \eqref{RDM} by tracing over the coordinates of second oscillator i.e. by setting $x_2'=x_2$ and computing ${\rho}(x_1,x_1',x_2)$. The reduced density matrix can then be evaluated as:
\be 
{\rho}(x_1,x_1')=\int_{-\infty}^{\infty}{\rho}(x_1,x_1',x_2)dx_2.
\ee
We mention the final form of the reduced density matrix which can be used to calculate the entanglement measures as:
  \begin{widetext}
	\be\label{e4} 
	\begin{matrix}
	\displaystyle {\rho}(x_1,x_1')=N\bigg(\frac{g_1g_2}{\pi}\bigg)\sqrt{\frac{\pi}{A}}\exp\biggr\{(2\lambda A_1+\lambda\alpha_0)+\biggr(\frac{-(P+Q)}{2}+\frac{(Q-P)^2}{4A}+\lambda\alpha_1\biggr)x_1^2\\ \displaystyle+\biggr(\frac{-(P^*+Q^*)}{2}+\frac{(Q^*-P^*)^2}{4A}+\lambda\alpha_2\biggr)x_1'^2\\ \displaystyle
	+\biggr(\frac{(Q-P)(Q^*-P^*)}{2A}+\lambda\alpha_3\biggr)x_1x_1'+\lambda\biggr[\alpha_4x_1^4+\alpha_5x_1'^4+\alpha_6x_1^3x_1'+\alpha_7x_1x_1'^3+\alpha_8x_1^2x_1'^2\biggr]\biggr\}.
	\end{matrix}~ \ee   
	\end{widetext} 
Here $N$ is the normalisation factor, while,
\be\label{coeff2}
\begin{aligned}
    A&=\biggr(\frac{g_1^2+g_2^2}{2}\biggr)\\
   B&=\biggr((Q-P)x_1+(Q^*-P^*)x_1'\biggr).
\end{aligned}
\ee
The values of coefficients, $\alpha_i$ for $i=0$ to $i=8$, are listed in appendix \ref{app:Table}

\subsection{Computing the value of $tr({{\rho}}^n)$}\label{trace}
In this subsection we outline analytical steps to compute the expression for $tr({{\rho}}^n)$ where $\rho$ is the reduced density matrix of Eq. \eqref{e4}. The reduced density matrix, given by equation (\ref{e4}) is clearly representing a non-Gaussian state. The calculation of entanglement entropy for such a non-Gaussian state is shown in \cite{2020JHEP...11..114C}. We follow a similar method and modify the same derivation to compute entropy for the state representing Eq. \eqref{e4}.\\
According to the definition of trace, considering that $x_{n+1}=x_1$, one can write:
\bea\label{product}
tr({{\rho}}^n)=\int dx_1dx_2\cdots dx_n {{\rho}}(x_1,x_2){{\rho}}(x_2,x_3)\cdots{{\rho}}(x_n,x_1),\nonumber
\\
\eea
where, ${{\rho}}(x_i,x_{i+1})$, for $i=1$ to $n$, represents the reduced density matrix given in equation (\ref{e4}). The product of density matrices in Eq. \eqref{product}, when evaluated gives:
\begin{widetext}
	\be\label{e5}{
	\begin{matrix}
\displaystyle 	tr({{\rho}}^n)=N^n\biggr(\frac{g_1g_2}{\pi}\biggr)^n\biggr(\sqrt{\frac{\pi}{A}}\biggr)^ne^{n(2\lambda A_1+\lambda\alpha_0)}\times\\\displaystyle \int d^nx\exp\biggr\{\biggr[\frac{-(P+P^*+Q+Q^*)}{2}+\frac{(Q-P)^2+(Q^*-P^*)^2}{4A}+\lambda(\alpha_1+\alpha_2)\biggr]\sum_{i=1}^{n}x_i^2\\\displaystyle+\biggr[\frac{(Q-P)(Q^*-P^*)}{2A}+\lambda\alpha_3\biggr]\sum_{i=1}^{n}x_ix_{i+1}\\\displaystyle+\lambda\biggr[(\alpha_4+\alpha_5)\sum_{i=1}^{n}x_i^4\\\displaystyle+(\alpha_6x_{i-1}^3+\alpha_7x_{i+1}^3)\sum_{i=1}^{n}x_i+\alpha_8\sum_{i=1}^{n}x_i^2\sum_{i=1}^{n}x_{i+1}^2\biggr]\biggr\}.
	\end{matrix}}~ \ee    
\end{widetext} 

We modify the above expression by using new coefficients $\beta_1,\beta_2...\beta_7$ tabulated in appendix \ref{app:Table}. Note that each $\beta_i$ for $i=1$ to $7$ is evaluated by substituting $P,Q,A$ and $B$ defined in Eq. \eqref{coeff1} and Eq. \eqref{coeff2}.
One can then show that,
\begin{widetext}
    \be\label{e7} {
	\begin{matrix}
\displaystyle	tr({{\rho}}^n)=N^n\biggr(\frac{g_1g_2}{\pi}\biggr)^n\biggr(\sqrt{\frac{\pi}{A}}\biggr)^ne^{n(2\lambda A_1+\lambda\alpha_0)}\int d^nx\biggr\{\exp\biggr[(\beta_1+\lambda\beta_2)\sum_{i=1}^{n}x_i^2+(\beta_3+\lambda\beta_4)\sum_{i=1}^{n}x_ix_{i+1}\biggr]\times \\
\displaystyle	 \exp\biggr\{\lambda[\beta_5\sum_{i=1}^{n}x_i^4+(\alpha_6x_{i-1}^3+\alpha_7x_{i+1}^3)\sum_{i=1}^{n}x_i+\beta_6\sum_{i=1}^{n}x_i^2\sum_{i=1}^{n}x_{i+1}^2]\biggr\}\biggr\}~.\end{matrix}}~ \ee  
\end{widetext} 

The integral in the above Eq. (\ref{e7}) can be solved using the steps shown in appendix \ref{trI}. \\
To evaluate normalisation factor $N$ one needs to set $tr({{\rho}}(x_1,x_1'))=1$ for the reduced density matrix in Eq. \eqref{e4}. Using $\mu$ and $\xi$ defined in appendix \ref{trI} the normalisation factor is given by,
\begin{widetext}
	\be\label{norm} {
	N^n=\biggr(\frac{g_1g_2}{\pi}\biggr)^{-n}\biggr(\sqrt{\frac{\pi}{A}}\biggr)^{-n}e^{-n(2\lambda A_1+\lambda\alpha_0)}\biggr(\frac{\beta}{\pi}\biggr)^{\frac{n}{2}}|1-\mu|^n\biggr(1-\frac{3\lambda(\beta_5+\beta_6+\beta_7)}{4\beta^2(1-\mu)^4}\biggr)^n}
	~. \ee   
\end{widetext}

Substituting (\ref{norm}) and Eq. \eqref{Guass} in Eq. \eqref{e110} from appendix \ref{trI}, one can obtain the final value of trace of $n^{th}$ order of the reduced density matrix of Eq. \eqref{e4}, as:
\begin{widetext}
	\be\label{trf} {
	tr({{\rho}}^n)=\frac{|1-\mu|^n}{|1-\mu^n|}\biggr(1-\frac{3\lambda(\beta_5+\beta_6+\beta_7)}{4\xi^2(1-\mu)^4}\biggr)^n\biggr\{1+n\lambda[3\beta_5(M^{-1}_{11})^2+\beta_6(M^{-1}_{11})^2+2(M^{-1}_{12})^2+3\beta_7(M^{-1}_{11}M^{-1}_{12})]\biggr\}}
	~. \ee   
\end{widetext}

\subsection{Entanglement Measures}\label{sec:EntM}
In this subsection we compute the two entanglement measures, viz., von Neumann entanglement entropy and Renyi entropy using the respective formulae for the reduced density matrix of Eq. \eqref{e4}.\\
The von Neumann entanglement entropy for a given density matrix ${{\rho}}$ is computed as,
$S_{VN}=-tr({{\rho}}\ln{{\rho}})$. As we know the explicit $n-$ dependence of $tr({{\rho}}^n)$ from Eq. \eqref{trf}, we instead use replica trick often given as \cite{1994NuPhB.424..443H},
\be\label{e13} 
S_{VN}=-\lim_{n \to 1} \frac{\partial}{\partial n} tr({{\rho}}^n).
\ee
Substituting the respective values of matrix inverses of, Eq. \eqref{MInv} in Eq. \eqref{trf} one can show that the von Neumann entropy is given by,
\begin{widetext}
	\be\label{e14} 
	     S_{VN}=-\frac{\mu\ln\mu+(1-\mu)\ln(1-\mu)}{(1-\mu)}+\lambda\biggr[\frac{3\mu\ln\mu}{\xi^2(\mu+1)(\mu-1)^5}\beta_5+\frac{\ln\mu(1+\mu+\mu^2)}{\xi^2(\mu+1)(\mu-1)^5}\beta_6+\frac{3\ln\mu(1+\mu)}{4\xi^2(\mu-1)^5}\beta_7\biggr]~.\ee
\end{widetext} 
Using Eq. \eqref{mu}, the above expression for von Neumann entropy can be approximated to first order in coupling constant $\lambda$ as,
\begin{widetext}
	\bea\label{vNf}
	     S_{VN}&=&-\frac{C_1 \ln C_1+(1-C_1)\ln(1-C_1)}{(1-C_1)}\nonumber\\
	     &&+\lambda\biggr[-\frac{C_1 C_2 \ln C_1}{(1-C_1)^2}-\frac{C_2 \ln C_1}{(1-C_1)}+\frac{3C_1 \ln C_1}{C_3^2(C_1+1)(C_1-1)^5}\beta_5+\frac{\ln C_1(1+C_1+C_1^2)}{C_3^2(C_1+1)(C_1-1)^5}\beta_6+\frac{3\ln C_1(1+C_1)}{4C_3^2(C_1-1)^5}\beta_7\biggr]\nonumber~.
	     \\\eea.
\end{widetext} 
The Renyi entropy of order $n$ can be evaluated using,
\be
S_R=\frac{1}{1-n}\ln[tr({{\rho}}^n)].
\ee
Substituting the respective values of matrix inverses, Eq. \eqref{MInv} in Eq. \eqref{trf} while using Eq. \eqref{mu} one can show that the Renyi entropy is given by,
\begin{widetext}
	\bea\label{Renyif}
	\begin{matrix}
	    \displaystyle S_{R}=\frac{1}{1-n}\biggr\{n\ln(1-C_2-\lambda C_3)-\ln(1-(C_2+\lambda)^n)\\
	    \displaystyle  +n\lambda\biggr[\biggr(-\frac{3}{4C_3^2(1-C_1)^4}+\frac{(C_1^{2n}-1)^2}{4C_3^2(1-C_1^n)^4(C_1^2-1)^2}\biggr)\beta_5\\
	   \displaystyle   +\biggr(-\frac{3}{4C_3^2(1-C_1)^4}+\frac{(C_1^{2n}-1)^2}{4C_3^2(1-C_1^n)^4(C_1^2-1)^2}+2\frac{(C_1^n+C_1^2)^2}{4C_1^2C_3^2(C_1^2-1)^2(C_1^n-1)^2}\biggr)\beta_6\\
	    \displaystyle  +\biggr(-\frac{3}{4C_3^2(1-C_1)^4}+\frac{3(C_1^{2n}-1)(C_1^n+C_1^2}{4C_1C_3^2(1-C_1^n)^2(C_1^2-1)^2(C_1^n-1)}\biggr)\beta_7\biggr]\biggr\}~.
	     \end{matrix} 
	 \eea
\end{widetext}
Note that the coefficients $C_i$ for $i=1,2,3,4$ in Eq. \eqref{vNf} and Eq. \eqref{Renyif} arise due to the analytical steps shown in appendix \ref{trI}. These coefficients are tabulated in appendix \ref{app:Table}.\\
Using the values of von Neumann entropy, Eq. \eqref{vNf} and Renyi entropy, Eq. \eqref{Renyif} one can verify, using first order approximation in coupling constant $\lambda$, 
\begin{equation*}
 \lim_{n \to 1}S_R=S_{VN}~.\end{equation*}
Note that the final formulae of von Neumann as well as Renyi entropies depend on $\beta_i$ for $i=1,2...7$. These coefficients $\beta_i$, given in  table of \ref{trace} depend on timescale $t$ and $\delta t$. Entanglement measures therefore depend on these timescales. We check this time-dependence by computing numerical values of both entanglement measures in section \ref{sec:numerical}.

\section{\textcolor{Sepia}{\textbf{ \large Numerical Results}}}
\label{sec:numerical}
\begin{figure*}[htb!]
	\centering

		\includegraphics[width=16cm,height=10cm]{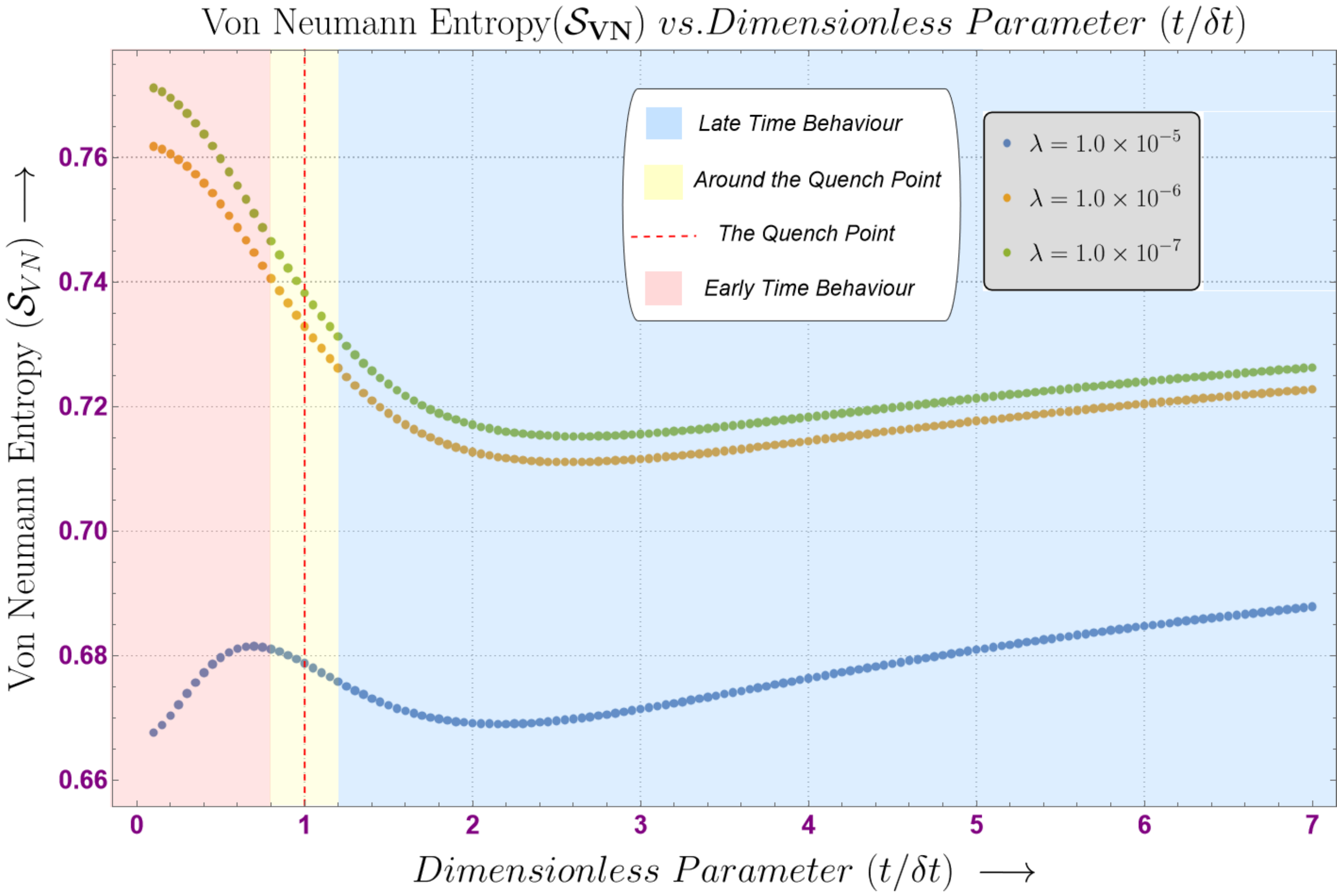}
	\caption{Variation of the Von Neumann entropy ($S_{VN})$ with respect to the dimensionless parameter ($t/\delta t)$ for different orders of the coupling constant $\lambda$ for {two coupled oscillators} with quartic perturbation.}
	\label{fig:vnl}
\end{figure*}

\begin{figure*}[htb!]
	\centering

		\includegraphics[width=16cm,height=10cm]{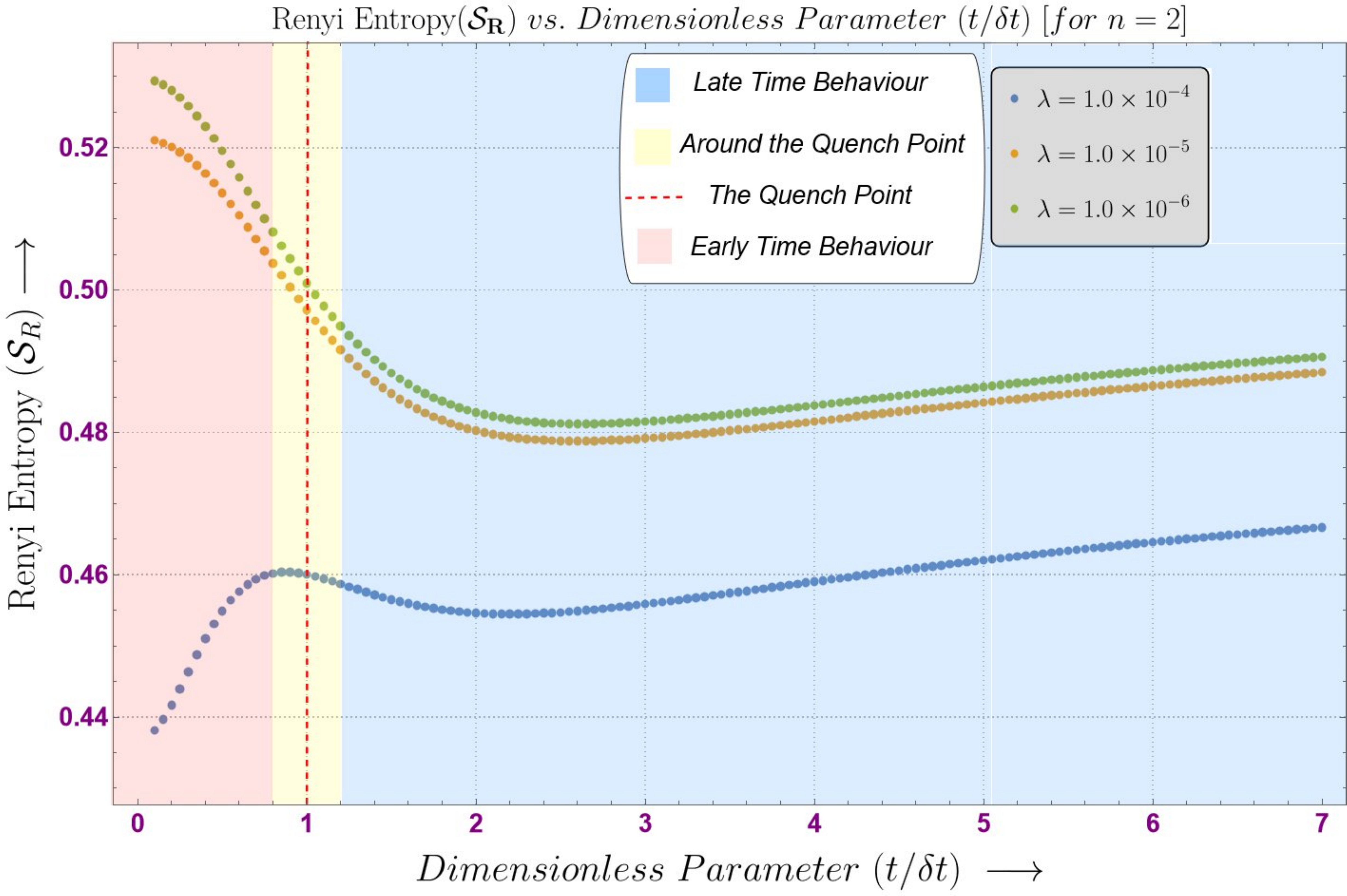}

	\caption{Variation of the Renyi entropy ($S_{R})$ for $n=2,$ with respect to the dimensionless parameter ($t/\delta t)$ for different orders of the coupling constant $\lambda$ for {two coupled oscillators} with quartic perturbation.}
	\label{fig:rnl}
\end{figure*}

\begin{figure*}[htb!]
	\centering

		\includegraphics[width=16cm,height=10cm]{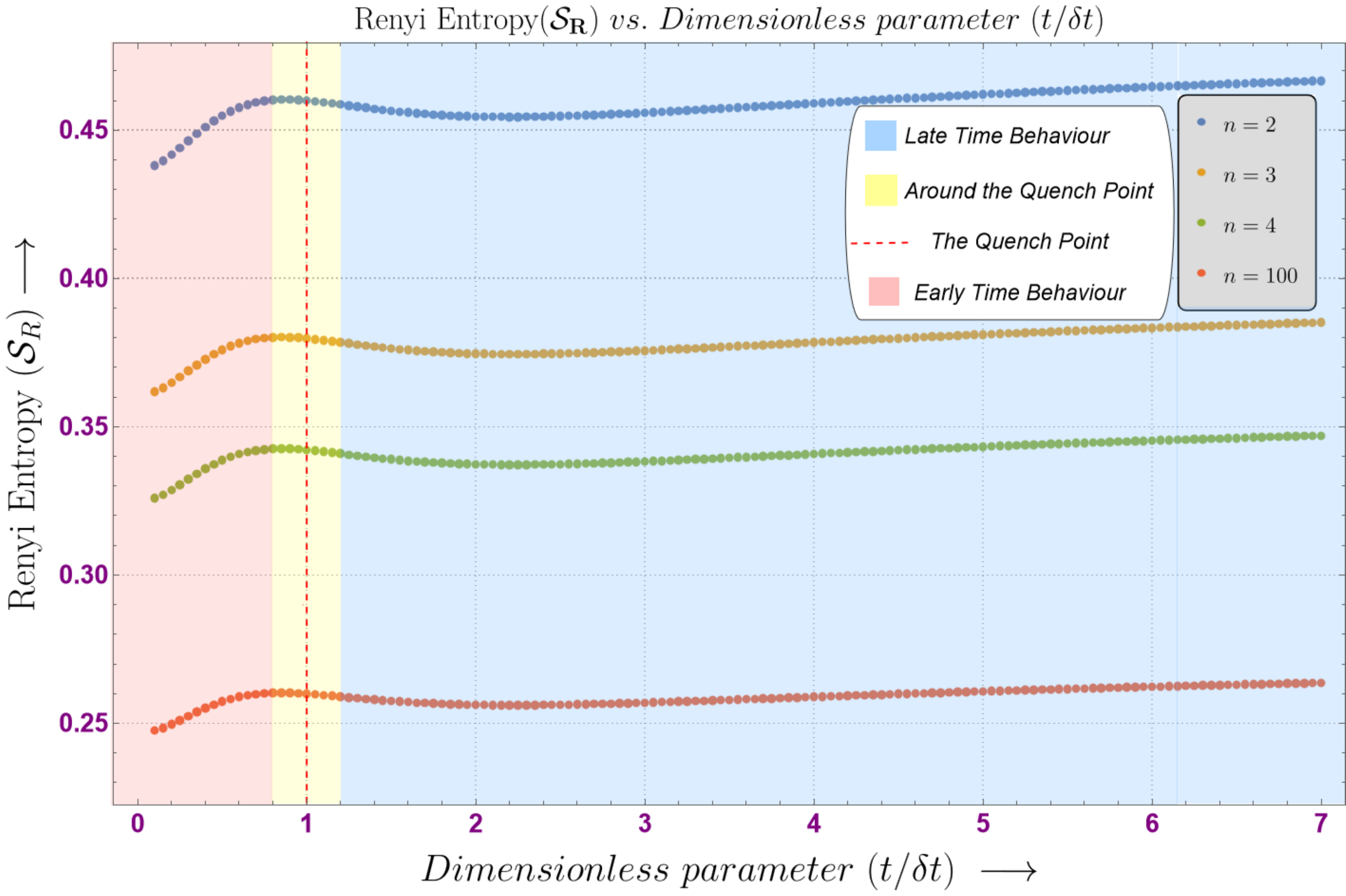}

	\caption{Variation of the Renyi entropy ($S_{R})$ for $\lambda=10^{-4},$ with respect to the dimensionless parameter ($t/\delta t)$ for different values of $n$ for {two coupled oscillators} with quartic perturbation.}
	\label{fig:rnl1}
\end{figure*}

In this section we numerically evaluate von Neumann and Renyi entanglement entropy measures computed for the quench setup of {two coupled oscillators} using Eq. \eqref{vNf} and Eq. \eqref{Renyif}. As mentioned before each factor in the derived formulae for entanglement measures  explicitly depends on $\sigma_i(t,\delta t)$ and $\gamma_i(t,\delta t)$. The values of $\sigma_i(t,\delta t)$ and $\gamma_i(t,\delta t)$ are computed by solving auxiliary equations, shown in, Eq. \eqref{aux} which is outlined in appendix \ref{MPequation}. Analytically solving the differential equation of Eq. \eqref{MPE} , is very complicated and hence we set some initial conditions to numerically evaluate the solution to this equation. \\
We begin by considering the coupling coefficient in the Hamiltonian of the coupled oscillators, Eq. \eqref{Eq_3.4} as $\eta=0.5$.
We further set the invariant quantities in Eq. \eqref{invar} as, $\Omega_1=\Omega_2=1$. To obtain the constants $A,B$ and $C$ mentioned in Eq. \eqref{expli} we first compute $\sigma_i(t,\delta t)$ and $\gamma_i(t,\delta t)$ at $t\rightarrow0$. Next we, set $d(t\rightarrow0)=f(t\rightarrow0)=0$, defined in, Eq. \eqref{d} and $\sigma_i(0,\delta t)=1$. Using these initial conditions we obtain values of $A,B$ and $C$ which can be inserted in Eq. \eqref{expli}. $\rho_i(t,\delta t)$ and $\gamma_i(t,\delta t)$ are then used to get numerical values of von Neumann and Renyi entanglement entropies, for the aforementioned initial conditions. \\
Using the numerical values of $S_{VN}$ and $S_R$, we parameterize four different plots for a chosen timescale. We have varied the dimensionless parameter $(t/\delta t)$ from $0.1$ to $0.7$ in steps of $0.05$. The ratio is then plotted on x-axis of respective figures. We term the value of $(t/\delta t)=1$ as the "Quench Point" represented by a red dotted line in all the respective figures. Using the values of $(t/\delta t)$ we divide all the plots in three different regions. The first region shaded as red, is marked for values of $(t/\delta t)<0.8$. This region shows the "early-time behavior" of the respective entanglement measures, when the quench rate $\delta t$ is varied in a way so as to keep $(t/\delta t)<<1$. The next region, shaded as yellow is marked by two equal intervals to the right as well as left of quench point, precisely for values of $(t/\delta t)$ between $0.8$ and $1.2$. This region represents the values of entanglement measures for $(t/\delta t)\approx1$ and hence is termed as the region "around the quench point". The last region shaded as blue is marked for values of  $(t/\delta t)>1.2$. This region shows the "late-time behavior" of the respective entanglement measures, when the quench rate $\delta t$ is varied in a way so as to keep $(t/\delta t)>>1$. \\

\begin{figure*}[htb!]
	\centering

		\includegraphics[width=16cm,height=10cm]{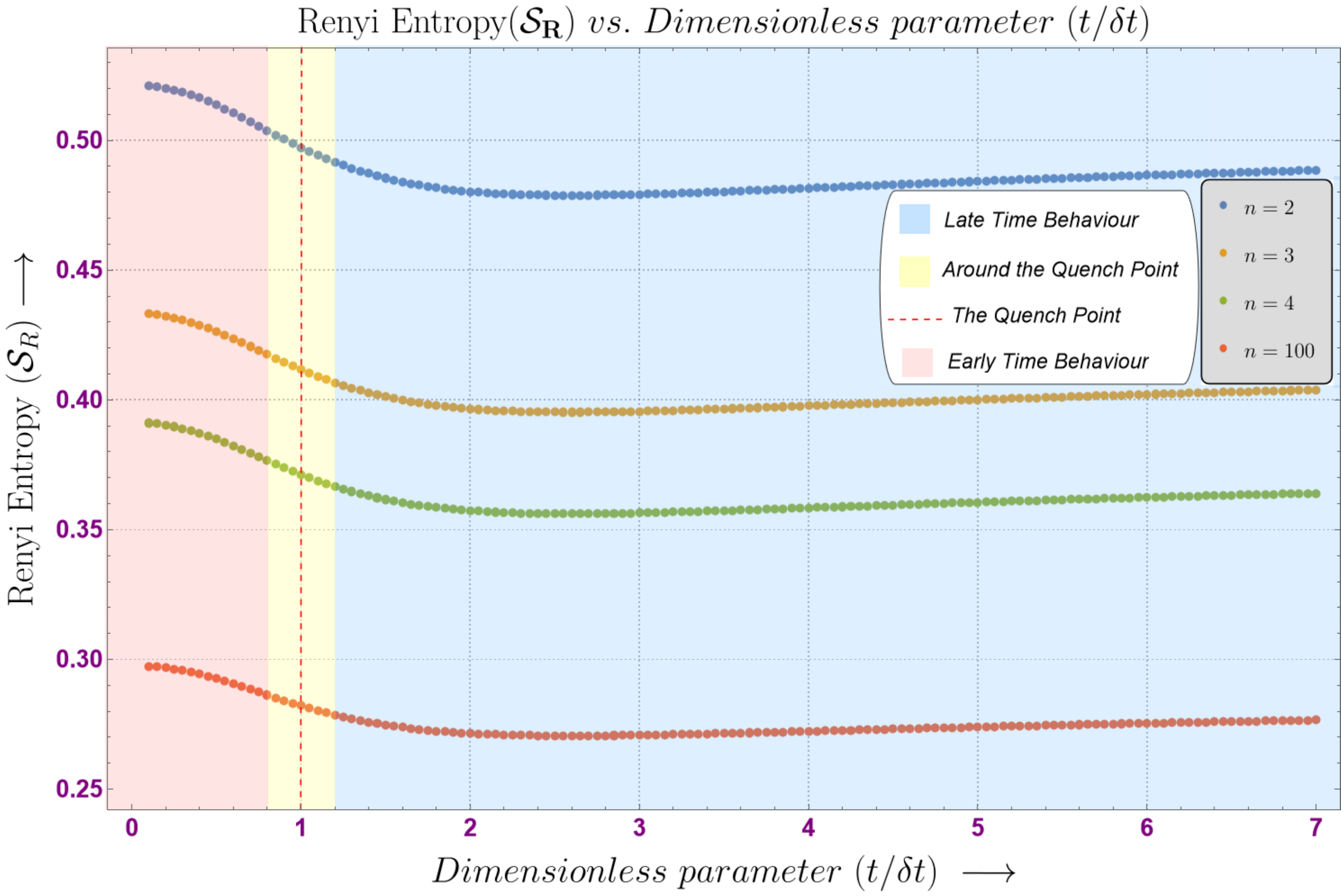}
		\caption{Variation of the Renyi entropy ($S_{R})$ for $\lambda=10^{-5},$ with respect to the dimensionless parameter ($t/\delta t)$ for different values of $n$ for {two coupled oscillators} with quartic perturbation.}
		\label{fig:rnl2}
\end{figure*}

 
In FIG. \ref{fig:vnl}. we have plotted the von Neumann (V-N) entropy for {two coupled oscillators having quartic self-coupling} with respect to the dimensionless parameter $(t/\delta t)$
for different orders of $\lambda$. We observe that the computed values of V-N entropy are negative for the chosen timescale for $\lambda>10^{-5}$. We begin by plotting the values of V-N entropy by decreasing the  order of  $\lambda$, starting from $\lambda=10^{-5}$. We see that for $\lambda=10^{-5}$ initially the V-N entropy grows for very small values of $(t/\delta t)<0.6$. It can therefore be inferred from the plot that V-N entropy (for $\lambda=10^{-5}$) increases in most of the region covering the early-time behaviour, shaded as blue. Further, in the range $0.6<(t/\delta t)<2.2$ the entropy decreases gradually. Thus, the whole region near to the Quench-Point, shaded as yellow, shows a decreasing value of V-N entropy. After $(t/\delta t)>2.2$ the V-N entropy increases monotonically and shows a thermalising behaviour for large values of $(t/\delta t)$. Most of the late-time behaviour therefore, shows thermalising behaviour of V-N entropy.\\
In case of $\lambda=10^{-6}$ and $\lambda=10^{-7}$ we observe from the plot in FIG. \ref{fig:vnl}, that there is a decrease in von Neumann entropy upto $(t/\delta t)<2.6$. Hence, the whole early-time behaviour region as well as the region around the Quench Point, shaded as red and yellow respectively, show a decreasing trend in V-N entropy. This trend is in contrast to the same for $\lambda=10^{-5}$.  When we move further towards larger values of the dimensionless parameter $(t/\delta t)$ we again see a thermalising behaviour for both the coupling constants. This region is shaded as blue and shows trend similar to that of $\lambda=10^{-5}$. Another observation which we can make from this graph is that as we decrease the order of the coupling constant $\lambda$ the von Neumann entropy increases.\\

In FIG. \ref{fig:rnl}. we have plotted the Renyi entropies for {two coupled oscillators having quartic self-coupling} with respect to the dimensionless parameter $(t/\delta t)$
for different orders of $\lambda$. We observe that the computed values of Renyi entropy are negative for the chosen timescale for $\lambda>10^{-4}$. We begin by plotting the values of Renyi entropy by decreasing the  order of  coupling constant, starting from $\lambda=10^{-4}$. We see that for $\lambda=10^{-4}$, initially the Renyi entropy grows for very small value of $(t/\delta t)<0.8$. Hence, the early time-behaviour of the system shows an increasing trend in values of Renyi entropy, shaded by red colour. Further, in the range $0.8<(t/\delta t)<2.2$ the entropy decreases gradually. The region around the Quench Point, shaded as yellow, shows decreasing trend in Renyi entropy. After $(t/\delta t)>2.2$ the Renyi entropy increases monotonically and shows a thermalising behaviour for large values of $(t/\delta t)$. Most of the late-time behaviour of the system shows the thermalisation trend in Renyi entropy. This region is shaded by blue colour. This behaviour is similar to that of V-N entropy for $\lambda=10^{-5}$ shown in FIG. {\ref{fig:vnl}}.\\
In case of $\lambda=10^{-5}$ and $\lambda=10^{-6}$ we observe from the graph that there is a decrease in Renyi entropy upto $(t/\delta t)<2.6$. Hence, the early-time behaviour as well as behaviour of the system around the Quench Point results in decreasing values of Renyi entropy, shaded as red and yellow respectively in the FIG. \ref{fig:rnl}. This trend is in contrast to that of $\lambda=10^{-4}$.  When we move further towards larger values of the dimensionless parameter $(t/\delta t)$ we again see a thermalising behaviour for both the coupling constants. The late-time behaviour of the system is, mostly characterised by thermalisation of Renyi entropy. This is shown by blue region in FIG. \ref{fig:rnl}. This behaviour is similar to that of V-N entropy for $\lambda=10^{-6},10^{-7}$ in FIG. \ref{fig:vnl}.   Another observation which we can make from this graph is that as we decrease the order of the coupling constant $\lambda$ the Renyi entropy increases.\\
Hence, as we are decreasing the order of $\lambda$ the plots of von Neumann entropy
and that of Renyi entropy show a similar behaviour with respect to each other, given that the order of $\lambda$ in the case of von Neumann entropy is 
one lower than that in Renyi entropy.\\

In the FIG. \ref{fig:rnl1}. we have plotted the Renyi entropies for {two coupled oscillators having quartic self-coupling} with respect to the dimensionless parameter $(t/\delta t)$
for different orders of Renyi entropy i.e. for different values of $n$, set at $\lambda=10^{-4}$. The early time behaviour shows an increasing trend in the value of Renyi entropy for the chosen values of $n=2,3,4$. The entropy then decreases covering the region around the quench. Most of the late time behaviour of the system shows thermalising nature of Renyi entropy. It is clear that this scaling behaviour is retained for large value of $n=100$.\\
FIG. \ref{fig:rnl2}. shows parametric variation for different orders of Renyi entropy i.e. for different values of $n$, set at $\lambda=10^{-5}$. The early time behaviour as well as the behaviour of system near the quench point, shows a decreasing trend in values of Renyi entropy. Most of the late-time behaviour is characterised again by thermalisation of Renyi entropy for chosen values of $n=2,3,4$. Again, this scaling behaviour is retained for large value of $n=100$.  
\section{\textcolor{Sepia}{\textbf{ \large Conclusion\label{sec:Conclusions}}}}
The concluding remarks of this work are appended below point-wise:
\begin{itemize} 
\item {Focusing on a system of two coupled oscillators with quartic perturbation, we have derived analytical expressions of von Neumann entanglement entropy and Renyi entropy, undergoing a quantum quench.} 
\item First we have computed the expression for eigenstates of unperturbed Hamiltonian using invariant operator method. Using this expression we have approximated the first order $\lambda\phi^4$ correction for the total Hamiltonian of the system. Since the Hamiltonian is time-dependent due to the chosen quench profile as the frequency of the oscillators, it is quite evident that the total Hamiltonian can be quantised by using solutions to the Ermakov-Milne-Pinney equation. The ground state of the total Hamiltonian of the system, having $\phi^4$ interaction term, is then used to derive analytical expressions for the respective entanglement measure.  
\item Next we have mentioned the reduced density matrix for the ground state of the above described system of coupled oscillators. This reduced density matrix, clearly represents non-Gaussian state due to presence of quartic interaction terms. We deal with this non-Gaussian terms by constructing a quartic tensor and computing the trace of $n^{th}$ order of reduced density matrix. 
\item Finally, we employ the use of replica trick for computing von Neumann entanglement entropy. Further, Renyi entropy was computed using the standard formula, depending on the reduced density matrix. The analytical expression for these entanglement measures is time-dependent as all the coefficients in the respective expression depend on solutions of Ermakov-Milne-Pinney equation.
\item Using the numerically evaluated values of von Neumann entropy and Renyi entropy we studied the variation of these entanglement measures with respect to the dimensionless parameter $(t/\delta t)$ specifying three regions: early-time behaviour, the behaviour around the quench point and the late-time behaviour. 
\item From these numerical results, we find that both von Neumann entropy and Renyi entropy delicately depend on the order of coupling constant $\lambda$. Evidently there exists a respective threshold order of $\lambda$ beyond which $\lambda$ if increased, we don't get positive values of both von Neumann and Renyi entanglement entropies, for chosen values of $(t/\delta t)$. For the respective threshold order of $\lambda$ we observe same scaling behavior in both von Neumann entropy and Renyi entropy. This scaling behavior can be characterised by a trend of increasing values of the entanglement measure for early times while in the region around the quench point the behavior shows a decreasing trend in these values. 
\item As the order of $\lambda$ is decreased below the respective threshold order we get another scaling behaviour of both von Neumann and Renyi entropies. This scaling behaviour can be characterised by a trend of decreasing values of the respective entanglement measure for both early-times and around the quench point regions. 
Both scaling behaviors show thermalising behaviour of the respective entanglement measures at very late times.
\item It is quite clear from the plots that as we decrease the order of $\lambda$ the value of both entanglement measures increases.
\item Also, for a given order of coupling constant $\lambda$ von Neumann entropy thermalises at higher values compared to that of Renyi entropy. 
\item Next we find that for a particular order of $\lambda$ we obtain same scaling behaviour for different orders of the Renyi entropy. However, the value of Renyi entropy decreases as we increase the order of Renyi entropy.
The particular scaling behaviour is retained even for the case of very high order of Renyi entropy.
\end{itemize}

\textbf{Future Prospects:}
\begin{itemize}
    
\item {In the present article, we have analyzed the effects of quantum quench on the entanglement entropy for a system consisting of two coupled oscillators with quartic perturbation. Of course, this study of entanglement entropy and quantum quench can be generalised to that of $N$-coupled oscillators. For $N\rightarrow\infty$, in the continuous limit, it would be interesting to explore the effects of quantum quench on entanglement in the context of interacting field theory.}
\item One of the latest developments in research in the field of high energy physics is, the study of circuit complexity \cite{cc1,cc2,cc3,cc4,cc5,cc6,cc7}. There are some works which are focussed on relating the complexity with quantum entanglement \cite{cent1,cent2,PhysRevLett.122.081601,cent3,cent4}. The study of the same might turn out to be intriguing in the case of interacting quenched field theories.
\item Hence, it would be interesting to explore the connection between quantum circuit complexity and entanglement and check its consistency with the CA and CV \cite{cacv1,cacv2,cacv3,cacv4} proposal.

\end{itemize}

{\bf Acknowledgement:} The Visiting Post Doctoral research fellowship of SC is supported by the J. C. Bose National Fellowship of Director, Professor Rajesh Gopakumar, ICTS, TIFR, Bengaluru. SC also would like to thank ICTS, TIFR, Bengaluru for providing the work friendly environment. SC also thanks all the members of our newly formed virtual international non-profit consortium Quantum Aspects of the Space-Time \& Matter (QASTM) for elaborative discussions.  RMG, SM,  NP, AR and PS would like to thank the members of the QASTM Forum for useful discussions. Last but not least, we would like to acknowledge our debt to the people belonging to the various parts of the world for their generous and steady support for research in natural sciences.

\onecolumngrid
\newpage
\appendix
\section{\textcolor{Sepia}{\textbf{ \large Computing Explicit Numerical values of $\sigma_i(T)$ and $\gamma_i(T)$\label{MPequation}}}}

Using auxiliary conditions given by equation (\ref{aux}) we briefly show the steps to compute $\sigma_1$ and $\gamma_1$. We begin by rearranging Eq. (\ref{aux}) for one of the oscillators,
\be\label{MPE}
\ddot{\sigma_1}+\omega_1^2(T)\sigma_1=\frac{\Omega_1}{\sigma_1^3}.\ee
The above second order differential equation is often termed as, Ermakov-Milne--Pinney equation \cite{ermakov2008second,milne1930numerical,pinney1950nonlinear}. This equation can be solved numerically to obtain $\sigma_1(T)$. Since $T=(t/\delta t)$, the solution will clearly be function of both $t$ and $\delta t$.
We assume that the form of solution of the above equation gives us a linear combination,
\be
\sigma_1(t,\delta t)=c_1z_1(t,\delta t)+c_2z_2(t,\delta t)~.
\ee
Here, $c_1$ and $c_2$ are numerical constants, while $z_1$ and $z_2$ are treated as two complex-valued solutions of Eq. \eqref{MPE}. We will consider only $z_1$ as one of the solutions. Using the form of quench profile Eq. \eqref{quenchprof}, the computed value of one of the solutions is,
\be
z_1(t,\delta t)=\left[e^{\frac{2 t}{\text{$\delta $t}}}\right]^{-\frac{1}{2} (i \text{$\delta $t})} \left[e^{\frac{2 t}{\text{$\delta $t}}}+1\right]^{\frac{1}{2} \left(\sqrt{1-4 \text{$\delta $t}^2}+1\right)} \, _2F_1\left[\frac{1}{2} \left(\sqrt{1-4 \text{$\delta $t}^2}+1\right),\frac{1}{2} \left(-2 i \text{$\delta $t}+\sqrt{1-4 \text{$\delta $t}^2}+1\right);1-i \text{$\delta $t};-e^{\frac{2 t}{\text{$\delta $t}}}\right]~.
\ee
Here, $_2F_1$ represents the hypergeometric function.
Since $z_1$ is complex valued we can write, $z_1=y_1+iy_2$ such that $y_1$ and $y_2$ are now two real-linearly independent equations. We give an outline of steps shown in \cite{2016arXiv160308747M} for numerical solution of Eq. \eqref{MPE} using these linearly independent equations.
This solution is guaranteed to be of the form,
\be\label{expli}
\sigma_1(t,\delta t)=[\sqrt{Ay_1^2(t,\delta t)t+2By_1(t,\delta t)y_2(t,\delta t)+Cy_2^2(t,\delta t)}]~.
\ee
The next step is to determine the constants in equation (\ref{expli}). These are fixed by setting the condition $AC-B^2=\Omega_1^2$. These steps give us the explicit value of $\sigma_1(t,\delta t)$. One can repeat these steps by inserting the respective parameters (of second oscillator) to find $\sigma_2(t,\delta t)$.\\
Since, $\Omega_i=\sigma_i^2\dot{\gamma_i}$, the explicit value of $\gamma_i$ is computed by using value of Eq. \eqref{expli},
\be
\gamma_i(t,\delta t)=\int^t_0\frac{\Omega_i}{\sigma_i^2(t,\delta t)}dt
\ee
Inserting the values of $\sigma_i(t,\delta t)$ and $\gamma_i(t,\delta t)$ it is clear that the wavefunction, in Eq. \eqref{wavef} becomes a function of both $t$ and $\delta t$. Note that we conceal this functional dependence in all the sections until the entropy is numerically evaluated in section \ref{sec:numerical}.

\section{\textcolor{Sepia}{\textbf{ \large An outline of Invariant operator representation\label{appendixA}}}}
In section \ref{sec:E2} we defined $I_j$ as an operator in Eq. \eqref{invar}. We outline in this appendix a few important steps for constructing this operator and the way to find eigenstates of this operator. Note that the subscript $j=1,2$ represent the parameters described for the oscillators having spatial coordinates: $X_1$ and $X_2$ respectively. \\
The operator $I_j$ is constructed such that it satisfies \cite{doi:10.1063/1.1664991},
\be
\frac{\partial I_j}{\partial T}+\frac{1}{i}[I_j,H_j],
\ee
where $H_j\hspace{2pt}\text{for}~j=1,2$ represents the respective decoupled Hamiltonian for each oscillator \eqref{decoup}. One can show that the operator given in Eq. \eqref{invar} satisfies the above condition. \\

It is assumed that the invariant $I_j$ is one of a complete set of commuting observables for respective $H_j$. This guarantees that there is a complete set of eigenstates for each $I_j$ defined in Eq. \eqref{invar}.
We refer to $u_{0_j}$ for $j=1,2$ as the ground state for the spectrum of the respective invariant operator. These ground states of the respective invariant operators  can be determined using the condition, $a_ju_{0_j}=0$ where $a_j$ is the respective  annihilation operator Eq. \eqref{crea}. When evaluated, the expression for ground state of the invariant operator $I_j$ is given by,
\bea
u_{0_j}=\biggr(\frac{\dot{\gamma}_j}{\pi}\biggr)^{1/4}\exp\biggr[-\frac{\dot{\gamma_j}}{2}(-i\dot{\sigma_j}\sigma_j\dot{\gamma_j})X_j^2\biggr].
\eea
Using the ground states and the respective creation operators $a_j^\dagger$ one can then show that, the $n^{th}$ eigenstate of the invariant-operator $I_j$ is given by,
\bea\label{eigenI}
u_{n_j}=\frac{1}{\sqrt{n!}}(a_j^\dagger)^nu_{0_j}=\biggr(\frac{1}{2^{n_j}n_j!}\biggr)\biggr(\frac{\dot{\gamma_j}}{\pi}\biggr)^{1/4}\exp\biggr[\dot{\gamma}_j\biggr(1-\frac{i\dot{\sigma_j}}{\sigma_j\dot{\gamma_j}}\biggr)X_j^2\biggr]\textbf{H}_{n_j}\left[\sqrt{\dot{\gamma_j}}X_j\right].
\eea
Here, $j=1,2$ and $\textbf{H}_{n_j}$ represents the Hermite polynomial of order $n_j$. Using the eigenstates of invariant operator \eqref{eigenI}, one can compute the wavefunctions of the decoupled Hamiltonians \cite{1994PhRvA..50.1035Y}. It can be shown that the computed wavefunctions take the form: $\psi_{n_j}=e^{i\alpha_{n_j}}u_{n_j}$, as solutions to Schrodinger's equations for respective $H_j$,
where $\alpha_{n_j}=-({1}/{2}+n_j)$; for $j=1,2$. The eigenstates for unperturbed Hamiltonian for the coupled oscillator system can further be computed as $\psi^{(0)}_{n_1,n_2}=\psi_{n_1}\times\psi_{n_2}$.\\

Using equation (\ref{eigenI}) one can then show that,
\begin{align}\label{eigen}
 {\psi_{n_1,n_2}^{(0)}}=&\sqrt{\frac{\dot{\gamma_1}^2\dot{\gamma_1}^2}{2^{n_1+n_2}n_1!n_2!\pi}}\exp\left[-i\frac{(2n_1+1)\gamma_1+(2n_2+1)\gamma_2}{2}\right]
\exp\left[-\frac{1}{2}\dot{\gamma_1}\Big(-\frac{i\dot{\sigma_1}}{\dot{\gamma_1}\sigma_1}\Big)X_1^2-\frac{1}{2}\dot{\gamma_2}\Big(-\frac{i\dot{\sigma_2}}{\dot{\gamma_2}\sigma_2}\Big)X_2^2\right]\times\nonumber\\ &\textbf{H}_{n_1}\left[\sqrt{\dot{\gamma_1}}X_1\right]\textbf{H}_{n_2}\left[\sqrt{\dot{\gamma_2}}X_2\right].
\end{align}

This equation \eqref{eigen} represents the eigenstates for the unperturbed Hamiltonian of two coupled oscillators having a quenched frequency profile. In \ref{perturbed} we compute the first order time-independent correction to the ground state of above equation.

\section{\textcolor{Sepia}{\textbf{ \large Computing Integrals in $tr({{\rho}}^n)$}\label{trI}}}
In the integral of equation (\ref{e7}) we have separated both Gaussian and non-Guassian parts. In this appendix we give detailed outline of solving both Gaussian and non-Gaussian contributions and finally combine them to compute the integral in (\ref{e7}). \\
The Gaussian part of the integrand can be parameterised by considering a quadratic coefficient matrix similar to the case in \cite{2020JHEP...11..114C}. Using the values of $\beta_i$ for $i=1$ to $i=7$ defined in section \ref{trace}, this coefficient matrix is defined as:
\be\label{e8}{
M_{ij}=-2(\beta_1+\lambda\beta_2)\delta_{ij}-(\beta_3+\lambda\beta_4)(\delta^i_{j+1}+\delta^{i+1}_j).}
~\ee
We further modify the above defined matrix by introducing two new symbols $\mu$ and $\xi$. We choose these variables so that they satisfy,
\bea\label{mu}
\xi(1+\mu^2)=-(\beta_1+\lambda\beta_2)\\
2\xi\mu=\beta_3+\lambda\beta_4~.\nonumber
\eea
We consider the following explicit solution of these equations, approximated to first order in $\lambda$:
\be\label{expl}
\begin{aligned}
\mu=C_1+\lambda C_2\\
\xi=C_3+\lambda C_4.
\end{aligned}
\ee
The values of newly defined coefficients $C_1,C_2...C_4$  are tabulated in appendix \ref{app:Table}.
Using matrix $M_{ij}$ defined in Eq. \eqref{e8}, one can recover the Gaussian part of Eq. \eqref{e7} as shown below,
\be\label{e9}
\exp\biggr[-\frac{1}{2}x^iM_{ij}x_j\biggr]=\exp\biggr[(\beta_1+\lambda\beta_2)\sum_{i=1}^{n}x_i^2+(\beta_3+\lambda\beta_4)\sum_{i=1}^{n}x_ix_{i+1}\biggr].
\ee
Moving on to the non-Guassian part in Eq. (\ref{e7}), we further define a quartic tensor as:
\bea\label{e101}
T_{ijkl}=\lambda\biggr[\beta_5\delta_{ij}\delta_{jk}\delta_{kl}+(\alpha_7\delta^{k-1}_l+\alpha_6\delta^{k+1}_l)\delta_{ij}\delta_{jk}+\beta_6\delta_{ij}\delta^{j+1}_k\delta_{kl}\biggr].
\eea
Using Eq. \eqref{e101}, one can recover the remaining part of Eq. \eqref{e7} as shown below,
\be\label{NG}
\exp\biggr[x^ix^jx^kx^lT_{ijkl}\biggr]=\exp\biggr\{\lambda\biggr[\beta_5\sum_{i=1}^{n}x_i^4+(\alpha_6x_{i-1}^3+\alpha_7x_{i+1}^3)\sum_{i=1}^{n}x_i+
\beta_6\sum_{i=1}^{n}x_i^2\sum_{i=1}^{n}x_{i+1}^2\biggr]\biggr\}.
\ee
Using the expression shown in Eq. \eqref{e9}  and the non-Gaussian contribution from Eq. \eqref{NG}, one can parametrise the integrand in Eq. (\ref{e7}),  as shown below:
\bea\label{e17} 
tr({{\rho}}^n)
  =N^n\biggr(\frac{g_1g_2}{\pi}\biggr)^n\biggr(\sqrt{\frac{\pi}{A}}\biggr)^ne^{n(2\lambda A_1+\lambda\alpha_0)}
  \int d^nx\exp\biggr[-\frac{1}{2}x^iM_{ij}x_j\biggr]
  \sum_{p=0}^{\infty}\frac{1}{i!}(x^ix^jx^kx^lT_{ijkl})^p.
~ \eea

Further, we define a Gaussian partition function, $Z_0$, given by,
\bea\label{Guass}
Z_0=\int d^nx \exp\biggr[-\frac{1}{2}x^iM_{ij}x_j\biggr]=\sqrt{\frac{(2\pi)^n}{\text{det}M}}=\frac{\left(\sqrt{\frac{\pi}{\xi}}\right)^n}{|1-\mu^n|},
\eea
where $\text{det}M$ denotes the determinant of matrix $M_{ij}$ of Eq. \eqref{e8}.
Using the above partition function the summed over tensor in Eq. \eqref{e17} can be transformed to a correlator as shown below retaining the form of perturbative expansion,

\be\label{corre}
   tr({{\rho}}^n)=N^n\biggr(\frac{g_1g_2}{\pi}\biggr)^n\biggr(\sqrt{\frac{\pi}{A}}\biggr)^ne^{n(2\lambda A_1+\lambda\alpha_0)}Z_0\sum_{p=0}^{\infty}\frac{1}{p!}\left\langle\left\langle x^{i_1}x^{j_1}x^{k_1}x^{l_1}...x^{i_p}x^{j_p}x^{k_p}x^{l_p}\right\rangle\right\rangle T_{i_1j_1k_1l_!}...T_{i_pj_pk_pl_p}.
  \ee

We simplify the above expression in Eq. \eqref{corre} using a generating functional $J$, as shown below,
\bea\label{GI}
Z(J)=\frac{1}{Z_0}\int d^nx\exp\biggr[-\frac{1}{2}x^iM_{ij}x_j+J_ix_i\biggr]
    =\exp\biggr[\frac{1}{2}J_iM^{-1}_{ij}J_j\biggr].
\eea
As shown in \cite{2020JHEP...11..114C}, correlator of Eq. \eqref{corre} computed using the above Eq. \eqref{GI} then becomes,
\be\label{e16}
\left\langle\left\langle x_{i_1}...x{i_{2m}}\right\rangle\right\rangle=\frac{\del^{2m}}{\del J_{i_1}...\del J_{i_{2m}}}Z(J)\biggr|_{J=0}
=\frac{1}{2^mm!}\sum_{\sigma\in S_{G}} \left(M^{-1}\right)_{i_{\sigma(1)\sigma(2)}}...\left(M^{-1}\right)_{i_{\sigma(2n-1)\sigma(2n)}}.
\ee
Here $G$ is the quotient group which can be defined to reduce the sum significantly. Note that for a $4m$ point correlator function the chosen quotient group gives rise to three different permutations. A more detailed discussion about finding the quotient group $G$ can be found in \cite{2020JHEP...11..114C}.\\
Using the value of quartic tensor from Eq. \eqref{e9} and the correlator from Eq. \eqref{e16}, one can simplify Eq. \eqref{corre} as,
\be\label{e110} {
	tr({{\rho}}^n)=N^nZ_0\biggr(\frac{g_1g_2}{\pi}\biggr)^n\biggr(\sqrt{\frac{\pi}{A}}\biggr)^ne^{n(2\lambda A_1+\lambda\alpha_0)}\left\{1+n\lambda\left[(3\beta_5+\beta_6)(M^{-1}_{11})^2+3\beta_7M^{-1}_{11}M^{-1}_{12}+2\beta_6\left(M^{-1}_{12}\right)^2\right]\right\}.
}~ \ee   
One can check that the matrix inverses are \cite{2020JHEP...11..114C},
\bea\label{MInv}
M^{-1}_{11}=\frac{(\mu^{2n}-1)}{2(1-\mu^n)^2\xi(\mu^2-1)};\hspace{1.5cm}
M^{-1}_{12}=\frac{(\mu^n+\mu^2)}{2\mu\xi(\mu^2-1)(\mu^n-1)}.
\eea
\section{Tabulated Values of Coefficients}\label{app:Table}
In this appendix, the values of various coefficients we have used in some steps to compute the analytical expression of entanglement measures, are tabulated in respective tables.\\
\begin{itemize}
\item We begin by listing the values of $A_i$ for $i=1,2...6$ in equation Eq. \eqref{wavef} of section \ref{perturbed} in the table given below.

\begin{longtable}{|c|c|}
\hline
\rowcolor{Gray}
 $A_{i}$ & Coefficient of $A_{i}$\\
\hline
$A_{1}$ &
{\Large \scalebox{0.70}{
\parbox[t]{20cm}
{\Large $\\\frac{3}{16} \biggr(-\frac{8 \sigma _1^2 \sigma _2^2}{\sigma _1^2 \omega _1^2+\sigma _2^2 {\omega_2}^2+\dot{\sigma }_1+\dot{\sigma }_2+\frac{1}{\sigma _1^2}+\frac{1}{\sigma _2^2}}+\frac{3 \sigma _1^4+4 \sigma _2^2 \sigma _1^2}{\sigma _1^2 \omega _1^2+\dot{\sigma }_1+\frac{1}{\sigma _1^2}}+\frac{3 \sigma _2^4+4 \sigma _1^2 \sigma _2^2}{\sigma _2^2\omega_2^2+\dot{\sigma }_2+\frac{1}{\sigma _2^2}}\biggr)
\\$}
}}\\ \hline
$A_{2}$ &
\scalebox{0.70}{
\parbox[t]{20cm}
 {\Large $\\\frac{3}{4} \sigma _1^2 \left(\frac{4 \sigma _2^2}{\sigma _1^2 \omega _1^2+\sigma _2^2 \omega_2^2+\dot{\sigma }_1+\dot{\sigma }_2+\frac{1}{\sigma _1^2}+\frac{1}{\sigma _2^2}}-\frac{\sigma _1^2+2 \sigma _2^2}{\sigma _1^2 \omega _1^2+\dot{\sigma }_1+\frac{1}{\sigma _2^2}}\right)
 \\$}
 }\\ \hline
$A_{3}$ &
\scalebox{0.70}{
\parbox[t]{20cm}
{\Large $\\\frac{3 \sigma _1^2 \sigma _2^2}{\sigma _1^2 \omega _1^2+\sigma _2^2\omega_2^2+\dot{\sigma }_1+\dot{\sigma }_2+\frac{1}{\sigma _1^2}+\frac{1}{\sigma _2^2}}-\frac{3 \left(\sigma _2^4+2 \sigma _1^2 \sigma _2^2\right)}{4 \left(\sigma _2^2\omega_2^2+\dot{\sigma }_2+\frac{1}{\sigma _2^2}\right)}
\\$}
 }\\ \hline

$A_{4}$ &
\scalebox{0.70}{
\parbox[t]{20cm}
 {\Large  $\\-\frac{\sigma _1^4}{4 \left(\sigma _1^2 \omega _1^2+\dot{\sigma }_1+\frac{1}{\sigma _1^2}\right)}
\\$}
 }\\ \hline

$A_{5}$ &
\scalebox{0.70}{
\parbox[t]{20cm}
 {\Large $\\-\frac{\sigma _2^4}{4 \left(\sigma _2^2 \omega _2^2+\dot{\sigma }_2+\frac{1}{\sigma _2^2}\right)};
\\
$}
 }\\ \hline

$A_{6}$ &
\scalebox{0.70}{
\parbox[t]{20cm}
 {\Large $\\-\frac{6 \sigma _1^2 \sigma _2^2}{\sigma _1^2 \omega _1^2+\sigma _2^2 \omega _2^2+\dot{\sigma }_1+\dot{\sigma }_2+\frac{1}{\sigma _1^2}+\frac{1}{\sigma _2^2}};
\\$}
 }\\  \hline\hline\hline

\end{longtable}

\item Next, we tabulate the values of coefficients $\alpha_i$ for $i=1,2...8$ in equation Eq. \eqref{e4} of section \ref{sub:RDM}.

\begin{longtable}{|c|c|}

\hline
\rowcolor{Gray}
 $\alpha_{i}$ & Coefficient of $\alpha_{i}$\\
\hline
$\alpha_{0}$ &

{\Large \scalebox{.70}{
\parbox[t]{20cm}
{ $\\\frac{(A_2+A_3)}{2A}+\frac{3(A_4+A_5+A_6)}{8A^2}\\$}}
 }\\ \hline
$\alpha_{1}$ &

{\Large \scalebox{0.70}{
\parbox[t]{20cm}
 {$\\\frac{(A_2+A_3)}{2}+\frac{(3A_4+3A_5-A_6)}{4A}+\frac{(A_2-A_3)(Q-P)}{2A}+\frac{(A_2+A_3)(Q-P)^2}{4A^2}+\frac{3(A_4-A_5)(Q-P)}{4A^2}+\frac{3(A_4+A_5+A_6)}{8A^3}(Q-P)^2
 \\$}}
 }\\ \hline
$\alpha_{2}$ &
\scalebox{0.70}{
\parbox[t]{20cm}
 {\Large $\\\frac{(A_2+A_3)}{2}+\frac{(3A_4+3A_5-A_6)}{4A}+\frac{(A_2-A_3)(Q^*-P^*)}{2A}+\frac{(A_2+A_3)(Q^*-P^*)^2}{4A^2}+\frac{3(A_4-A_5)(Q^*-P^*)}{4A^2}+\frac{3(A_4+A_5+A_6)}{8A^3}(Q^*-P^*)^2
 \\$}
 }\\ \hline
$\alpha_{3}$ &
\scalebox{0.70}{
\parbox[t]{20cm}
{\Large $\\\frac{(A_2-A_3)}{2A}(Q+Q^*-P-P^*)+\frac{(A_2+A_3)(Q-P)(Q^*-P^*)}{2A^2}+\frac{3(A_4-A_5)(Q+Q^*-P-P^*)}{4A_2}+\frac{3(A_4+A_5+A_6)(Q-P)(Q^*-P^*)}{4A^3}
\\$}
 }\\ \hline
$\alpha_{4}$ &
\scalebox{0.70}{
\parbox[t]{20cm}
 {\Large  $\\\frac{(A_4+A_5+A_6)}{4}+\frac{(A_4-A_5)(Q-P)}{2A}+\frac{(3A_4+A_5-A_6)(Q-P)^2}{8A^2}+\frac{(A_4-A_5)(Q-P)^3}{8A^3}+\frac{(A_4+A_5+A_6)(Q-P)^4}{32A^4}
\\$}
 }\\ \hline
$\alpha_{5}$ &
\scalebox{0.70}{
\parbox[t]{20cm}
 {\Large $\\\frac{(A_4+A_5+A_6)}{4}+\frac{(A_4-A_5)(Q^*-P^*)}{2A}+\frac{(3A_4+A_5-A_6)(Q^*-P^*)^2}{8A^2}+\frac{(A_4-A_5)(Q^*-P^*)^3}{8A^3}+\frac{(A_4+A_5+A_6)(Q^*-P^*)^4}{32A^4}
 \\
$}
 }\\ \hline
$\alpha_{6}$ &
\scalebox{0.70}{
\parbox[t]{20cm}
 {\Large $\\\frac{(A_4-A_5)(Q^*-P^*)}{2A}+\frac{(3A_4+3A_5-A_6)(Q-P)(Q^*-p^*)}{4A^2}+\frac{(A_4-A_5)[3(Q-P)^2(Q^*-P^*)+(Q-P)^3}{8A^3}+\frac{(A_4+A_5+A_6)(Q-P)^3(Q^*-P^*)}{8A^4}
\\$}
 }\\ \hline  
$\alpha_{7}$ &
\scalebox{0.70}{
\parbox[t]{20cm}
 {\Large $\\\frac{(A_4-A_5)(Q-P)}{2A}+\frac{(3A_4+3A_5-A_6)(Q-P)(Q^*-p^*)}{4A^2}+\frac{(A_4-A_5)[3(Q-P)(Q^*-P^*)^2+(Q^*-P^*)^3}{8A^3}+\frac{(A_4+A_5+A_6)(Q-P)(Q^*-P^*)^3}{8A^4}
\\$}
 }\\ \hline  
$\alpha_{8}$ &
\scalebox{0.70}{
\parbox[t]{20cm}
 {\Large $\\\frac{(3A_4+3A_5-A_6)[(Q^*-P^*)^2+(Q-P)^2]}{8A^2}+\frac{(A_4-A_5)[3(Q-P)^2(Q^*-P^*)+3(Q-P)(Q^*-P^*)^2]}{8A^3}+\frac{(A_4+A_5+A_6)6(Q-P)^2(Q^*-P^*)^2}{32A^4}
\\$}
 }\\ \hline\hline \hline
 \end{longtable}

\item The values of $\beta_i$ for $i=1,2...7$ in Eq. \eqref{e7} of section \ref{trace} are tabulated in the below given table. 
 
 \begin{longtable}{|c|c|}
\hline
\rowcolor{Gray}
 $\beta_{i}$ & Coefficient of $\beta_{i}$\\
\hline
$\beta_{1}$ &
{\Large \scalebox{.70}{
\parbox[t]{20cm}
{ $\\-A+\frac{(g_2^2-g_1^2)^2-(dg_1^2-fg_2^2)^2}{8A}
\\$}}
 }\\ \hline
$\beta_{2}$ &
\scalebox{0.70}{
\parbox[t]{20cm}
 {\Large $\\A_2+A_3+\frac{3A_4+3A_5-A_6}{2A}+\biggr(\frac{A_2-A_3}{2A}\biggr)(g_2^2-g_1^2)+\frac{3(A_4-A_5)(g_2^2-g_1^2)}{4A^2}+[\frac{3(A_4+A_5+A_6)}{16A^3}+\frac{A_2+A_3}{8A^2}][(g_2^2-g_1^2)^2-(dg_1^2-fg_2^2)^2]
 \\$}
 }\\ \hline
$\beta_{3}$ &
\scalebox{0.70}{
\parbox[t]{20cm}
{\Large $\\\frac{(g_2^2-g_1^2)^2+(dg_1^2-fg_2^2)^2}{8A}
\\$}
 }\\ \hline
$\beta_{4}$ &
\scalebox{0.70}{
\parbox[t]{20cm}
 {\Large  $\\\biggr[\frac{A_2-A_3}{2A}+\frac{3(A_4-A_5)}{4A^2}\biggr][g_2^2-g_1^2]+\biggr[\frac{(A_2+A_3)}{8A^2}+\frac{3(A_4+A_5+A_6)}{16A^3}\biggr][(g_2^2-g_1^2)^2+(dg_1^2-fg_2^2)^2]
\\$}
 }\\ \hline
$\beta_{5}$ &
\scalebox{0.70}{
\parbox[t]{20cm}
 {\Large $\\\frac{A_4+A_5+A_6}{2}+\frac{(A_4-A_5)(g_2^2-g_1^2)}{2A}+\frac{3A_4+3A_5-A_6}{16A^2}[(g_2^2-g_1^2)^2-(dg_1^2-fg_2^2)^2]+\frac{A_4-A_5}{64A^3}[2(g_2^2-g_1^2)^3-6(g_2^2-g_1^2)(dg_1^2-fg_2^2)^2]+\frac{A_4+A_5+A_6}{512A^4}[2(g_2^2-g_1^2)^4+2(dg_1^2-fg_2^2)^4-12(g_2^2-g_1^2)^2(dg_1^2-fg_2^2)^2]
\\$}
 }\\ \hline
$\beta_{6}$ &
\scalebox{0.70}{
\parbox[t]{20cm}
 {\Large $\\\biggr(\frac{3A_4+3A_5-A_6}{16A^2}\biggr)[(g_2^2-g_1^2)^2-(dg_1^2-fg_2^2)^2]+\biggr(\frac{3(A_4-A_5)}{32A^3}\biggr)[(g_2^2-g_1^2)^3+(g_2^2-g_1^2)(dg_1^2-fg_2^2)^2]+\biggr(\frac{6(A_4+A_5+A_6)}{512A^4}\biggr)[(g_2^2-g_1^2)^4+(dg_1^2-fg_2^2)^4+2(g_2^2-g_1^2)^2(dg_1^2-fg_2^2)^2]
\\$}
 }\\ \hline
$\beta_{7}$ &
\scalebox{0.70}{
\parbox[t]{20cm}
 {\Large $\\\frac{(A_4-A_5)(g_2^2-g_1^2)}{2A}+\biggr(\frac{3A_4+3A_5-A_6}{8A^2}\biggr)[(g_2^2-g_1^2)^2+(dg_1^2-fg_2^2)]+\biggr(\frac{A_4-A_5}{32A^3}\biggr)[(g_2^2-g_1^2)^3]+\biggr(\frac{A_4+A_5+A_6}{64A^4}\biggr)[(g_2^2-g_1^2)^4-(dg_1^2-fg_2^2)^4]
\\$}
 }\\  \hline\hline\hline
\end{longtable}

\item The last table, given below contains values of $C_i$ for $i=1,2,3,4$ in Eq. \eqref{expl} of section \ref{trI}.

\begin{longtable}{|c|c|}

\hline
\rowcolor{Gray}
 $C_{i}$ & Value of $C_{i}$\\
\hline
$C_{1}$ &

{\Large \scalebox{.70}{
\parbox[t]{20cm}
{ $\\-\frac{\beta_1}{\beta_3}+\frac{(\beta_1^2-\beta_3^2)^{\frac{1}{2}}}{\beta_3}\\$}}
 }\\ \hline

$C_{2}$ &
\scalebox{0.70}{
\parbox[t]{20cm}
 {\Large $\\-\frac{\beta_2}{\beta_3}+\frac{(\beta_1\beta_2-\beta_3\beta_4)}{\beta_3(\beta_1^2-\beta_3^2)^{\frac{1}{2}}}+\frac{\beta_1\beta_4}{\beta_3^2}-\frac{\beta_4(\beta_1^2-\beta_3^2)^{\frac{1}{2}}}{\beta_3^2}
\\$}
 }\\ \hline
$C_{3}$ &
\scalebox{0.70}{
\parbox[t]{20cm}
{\Large $\\\frac{\beta_3}{2C_1}
\\$}
 }\\ \hline
$C_{4}$ &
\scalebox{0.70}{
\parbox[t]{20cm}
 {\Large  $\\\frac{\beta_4}{2C_1}-\frac{C_2\beta_3}{2C_1^2}
\\$}
 }\\ \hline\hline\hline
\end{longtable}

\twocolumngrid

\end{itemize}

\bibliography{referencesnew}

\providecommand{\href}[2]{#2}\begingroup\raggedright\begin{thebibliography}{10}

\bibitem{intro:merge1}
P.~{Calabrese} and J.~{Cardy}, ``{Entanglement entropy and quantum field
  theory},'' \href{http://dx.doi.org/10.1088/1742-5468/2004/06/P06002}{{\em
  Journal of Statistical Mechanics: Theory and Experiment} {\bfseries 2004}
  no.~6, (June, 2004) 06002},
  \href{http://arxiv.org/abs/hep-th/0405152}{{\ttfamily arXiv:hep-th/0405152
  [hep-th]}}.

\bibitem{intro:merge3}
E.~{Witten}, ``{Notes on Some Entanglement Properties of Quantum Field
  Theory},'' {\em arXiv e-prints} (Mar., 2018) arXiv:1803.04993,
  \href{http://arxiv.org/abs/1803.04993}{{\ttfamily arXiv:1803.04993
  [hep-th]}}.

\bibitem{intro:merge4}
T.~{Nishioka}, ``{Entanglement entropy: Holography and renormalization
  group},'' \href{http://dx.doi.org/10.1103/RevModPhys.90.035007}{{\em Reviews
  of Modern Physics} {\bfseries 90} no.~3, (July, 2018) 035007},
  \href{http://arxiv.org/abs/1801.10352}{{\ttfamily arXiv:1801.10352
  [hep-th]}}.

\bibitem{intro:merge5}
D.~{Blanco}, ``{Quantum information measures and their applications in quantum
  field theory},'' {\em arXiv e-prints} (Feb., 2017) arXiv:1702.07384,
  \href{http://arxiv.org/abs/1702.07384}{{\ttfamily arXiv:1702.07384
  [hep-th]}}.

\bibitem{intro:merge6}
M.~{Headrick}, ``{Lectures on entanglement entropy in field theory and
  holography},'' {\em arXiv e-prints} (July, 2019) arXiv:1907.08126,
  \href{http://arxiv.org/abs/1907.08126}{{\ttfamily arXiv:1907.08126
  [hep-th]}}.

\bibitem{intro:merge7}
L.~{Amico}, R.~{Fazio}, A.~{Osterloh}, and V.~{Vedral}, ``{Entanglement in
  many-body systems},'' \href{http://dx.doi.org/10.1103/RevModPhys.80.517}{{\em
  Reviews of Modern Physics} {\bfseries 80} no.~2, (Apr., 2008) 517--576},
  \href{http://arxiv.org/abs/quant-ph/0703044}{{\ttfamily
  arXiv:quant-ph/0703044 [quant-ph]}}.

\bibitem{intro:merge8}
J.~I. {Cirac}, ``{Entanglement in many-body quantum systems},'' {\em arXiv
  e-prints} (May, 2012) arXiv:1205.3742,
  \href{http://arxiv.org/abs/1205.3742}{{\ttfamily arXiv:1205.3742
  [quant-ph]}}.

\bibitem{intro:time-ent1}
A.~Lakshminarayan and V.~Subrahmanyam, ``Multipartite entanglement in a
  one-dimensional time-dependent ising model,''
  \href{http://dx.doi.org/10.1103/PhysRevA.71.062334}{{\em Phys. Rev. A}
  {\bfseries 71} (Jun, 2005) 062334}.

\bibitem{intro:time-ent2}
H.~F. Song, S.~Rachel, C.~Flindt, I.~Klich, N.~Laflorencie, and K.~Le~Hur,
  ``Bipartite fluctuations as a probe of many-body entanglement,''
  \href{http://dx.doi.org/10.1103/PhysRevB.85.035409}{{\em Phys. Rev. B}
  {\bfseries 85} (Jan, 2012) 035409}.

\bibitem{intro:time-ent3}
C.~J. Bardeen, ``Time dependent correlations of entangled states with
  nondegenerate branches and possible experimental realization using singlet
  fission,'' \href{http://dx.doi.org/10.1063/1.5117155}{{\em The Journal of
  Chemical Physics} {\bfseries 151} no.~12, (2019) 124503}.

\bibitem{intro:time-ent4}
A.~Sivaramakrishnan, ``Entanglement entropy with a time-dependent
  hamiltonian,'' \href{http://dx.doi.org/10.1103/PhysRevD.97.066003}{{\em Phys.
  Rev. D} {\bfseries 97} (Mar, 2018) 066003}.

\bibitem{intro:time-ent5}
P.~{Caputa}, G.~{Mandal}, and R.~{Sinha}, ``{Dynamical entanglement entropy
  with angular momentum and U(1) charge},''
  \href{http://dx.doi.org/10.1007/JHEP11(2013)052}{{\em Journal of High Energy
  Physics} {\bfseries 2013} (Nov., 2013) 52},
  \href{http://arxiv.org/abs/1306.4974}{{\ttfamily arXiv:1306.4974 [hep-th]}}.

\bibitem{intro:time-ent7}
E.~{Canovi}, E.~{Ercolessi}, P.~{Naldesi}, L.~{Taddia}, and D.~{Vodola},
  ``{Dynamics of entanglement entropy and entanglement spectrum crossing a
  quantum phase transition},''
  \href{http://dx.doi.org/10.1103/PhysRevB.89.104303}{{\em \prb} {\bfseries 89}
  no.~10, (Mar., 2014) 104303},
  \href{http://arxiv.org/abs/1311.3612}{{\ttfamily arXiv:1311.3612
  [cond-mat.stat-mech]}}.

\bibitem{intro:time-ent8}
P.~Jacquod and C.~Petitjean, ``Decoherence, entanglement and irreversibility in
  quantum dynamical systems with few degrees of freedom,''
  \href{http://dx.doi.org/10.1080/00018730902831009}{{\em Advances in Physics}
  {\bfseries 58} no.~2, (2009) 67--196}.

\bibitem{ent1}
S.~Akhtar, S.~Choudhury, S.~Chowdhury, D.~Goswami, S.~Panda, and A.~Swain,
  ``{Open Quantum Entanglement: A study of two atomic system in static patch of
  de Sitter space},''
  \href{http://dx.doi.org/10.1140/epjc/s10052-020-8302-2}{{\em Eur. Phys. J. C}
  {\bfseries 80} no.~8, (2020) 748},
  \href{http://arxiv.org/abs/1908.09929}{{\ttfamily arXiv:1908.09929
  [hep-th]}}.

\bibitem{ent2}
S.~Choudhury and S.~Panda, ``{Quantum entanglement in de Sitter space from
  stringy axion: An analysis using $\alpha$ vacua},''
  \href{http://dx.doi.org/10.1016/j.nuclphysb.2019.03.018}{{\em Nucl. Phys. B}
  {\bfseries 943} (2019) 114606},
  \href{http://arxiv.org/abs/1712.08299}{{\ttfamily arXiv:1712.08299
  [hep-th]}}.

\bibitem{ent3}
S.~Choudhury and S.~Panda, ``{Entangled de Sitter from stringy axionic Bell
  pair I: an analysis using Bunch\textendash{}Davies vacuum},''
  \href{http://dx.doi.org/10.1140/epjc/s10052-017-5503-4}{{\em Eur. Phys. J. C}
  {\bfseries 78} no.~1, (2018) 52},
  \href{http://arxiv.org/abs/1708.02265}{{\ttfamily arXiv:1708.02265
  [hep-th]}}.

\bibitem{doi:10.1063/1.1664991}
H.~R. Lewis and W.~B. Riesenfeld, ``An exact quantum theory of the
  time‐dependent harmonic oscillator and of a charged particle in a
  time‐dependent electromagnetic field,''
  \href{http://dx.doi.org/10.1063/1.1664991}{{\em Journal of Mathematical
  Physics} {\bfseries 10} no.~8, (1969) 1458--1473}.

\bibitem{intro:invar1}
K.~Andrzejewski, ``{Dynamics of entropy and information of time-dependent
  quantum systems: exact results},''
  \href{http://dx.doi.org/10.1007/s11128-022-03440-w}{{\em Quant. Inf. Proc.}
  {\bfseries 21} no.~3, (2022) 117},
  \href{http://arxiv.org/abs/2108.00975}{{\ttfamily arXiv:2108.00975
  [quant-ph]}}.

\bibitem{intro:invar2}
B.~{Khantoul}, A.~{Bounames}, and M.~{Maamache}, ``{On the invariant method for
  the time-dependent non-Hermitian Hamiltonians},''
  \href{http://dx.doi.org/10.1140/epjp/i2017-11524-7}{{\em European Physical
  Journal Plus} {\bfseries 132} no.~6, (June, 2017) 258},
  \href{http://arxiv.org/abs/1610.09273}{{\ttfamily arXiv:1610.09273
  [quant-ph]}}.

\bibitem{invarSC}
S.~Choudhury, ``{Cosmological Geometric Phase From Pure Quantum States: A study
  without/with having Bell's inequality violation},''
  \href{http://arxiv.org/abs/2105.06254}{{\ttfamily arXiv:2105.06254 [gr-qc]}}.

\bibitem{intro:invar3}
J.-R. Choi, ``Coherent and squeezed states for light in homogeneous conducting
  linear media by an invariant operator method,'' {\em International Journal of
  Theoretical Physics} {\bfseries 43} no.~10, (2004) 2113--2136.

\bibitem{intro:invar4}
H.~Kanasugi and H.~Okada, ``{Systematic Treatment of General Time-Dependent
  Harmonic Oscillator in Classical and Quantum Mechanics},''
  \href{http://dx.doi.org/10.1143/ptp/93.5.949}{{\em Progress of Theoretical
  Physics} {\bfseries 93} no.~5, (05, 1995) 949--960}.

\bibitem{ermakov2008second}
V.~P. Ermakov, ``Second-order differential equations: conditions of complete
  integrability,'' {\em Applicable Analysis and Discrete Mathematics} (2008)
  123--145.

\bibitem{milne1930numerical}
W.~Milne, ``The numerical determination of characteristic numbers,'' {\em
  Physical Review} {\bfseries 35} no.~7, (1930) 863.

\bibitem{pinney1950nonlinear}
E.~Pinney, ``The nonlinear differential equation y+ p (x) y+ cy- 3= 0,'' in
  {\em Proc. Amer. Math. Soc}, vol.~1, pp.~681--681.
\newblock 1950.

\bibitem{2020FrP.....8..189C}
J.~R. {Choi}, ``{Perturbation theory for time-dependent quantum systems
  involving complex potentials},''
  \href{http://dx.doi.org/10.3389/fphy.2020.00189}{{\em Frontiers in Physics}
  {\bfseries 8} (June, 2020) 189}.

\bibitem{intro:pert-adiab1}
``Adiabatic evolution under quantum control,''
  \href{http://dx.doi.org/https://doi.org/10.1016/j.aop.2012.01.001}{{\em
  Annals of Physics} {\bfseries 327} no.~5, (2012) 1293--1303}.

\bibitem{intro:pert-adiab2}
M.-Y. Ye, X.-F. Zhou, Y.-S. Zhang, and G.-C. Guo, ``Two kinds of quantum
  adiabatic approximation,''
  \href{http://dx.doi.org/https://doi.org/10.1016/j.physleta.2007.03.056}{{\em
  Physics Letters A} {\bfseries 368} no.~1, (2007) 18--24}.

\bibitem{intro:Gaussianent2}
K.~Audenaert, J.~Eisert, M.~B. Plenio, and R.~F. Werner, ``Entanglement
  properties of the harmonic chain,''
  \href{http://dx.doi.org/10.1103/PhysRevA.66.042327}{{\em Phys. Rev. A}
  {\bfseries 66} (Oct, 2002) 042327}.

\bibitem{intro:Gaussianent3}
A.~{Jafarizadeh} and M.~A. {Rajabpour}, ``{Bipartite entanglement entropy of
  the excited states of free fermions and harmonic oscillators},''
  \href{http://dx.doi.org/10.1103/PhysRevB.100.165135}{{\em \prb} {\bfseries
  100} no.~16, (Oct., 2019) 165135},
  \href{http://arxiv.org/abs/1907.09806}{{\ttfamily arXiv:1907.09806
  [cond-mat.str-el]}}.

\bibitem{Gaussianent4}
T.~J. Osborne and M.~A. Nielsen, ``Entanglement in a simple quantum phase
  transition,'' \href{http://dx.doi.org/10.1103/PhysRevA.66.032110}{{\em Phys.
  Rev. A} {\bfseries 66} (Sep, 2002) 032110}.

\bibitem{intro:Gaussianent1}
R.~D. {Sorkin}, ``{Expressing entropy globally in terms of (4D)
  field-correlations},'' {\em arXiv e-prints} (May, 2012) arXiv:1205.2953,
  \href{http://arxiv.org/abs/1205.2953}{{\ttfamily arXiv:1205.2953 [hep-th]}}.

\bibitem{1994NuPhB.424..443H}
C.~{Holzhey}, F.~{Larsen}, and F.~{Wilczek}, ``{Geometric and renormalized
  entropy in conformal field theory},''
  \href{http://dx.doi.org/10.1016/0550-3213(94)90402-2}{{\em Nuclear Physics B}
  {\bfseries 424} no.~3, (Aug., 1994) 443--467},
  \href{http://arxiv.org/abs/hep-th/9403108}{{\ttfamily arXiv:hep-th/9403108
  [hep-th]}}.

\bibitem{2020JHEP...11..114C}
Y.~{Chen}, L.~{Hackl}, R.~{Kunjwal}, H.~{Moradi}, Y.~K. {Yazdi}, and
  M.~{Zilh{\~a}o}, ``{Towards spacetime entanglement entropy for interacting
  theories},'' \href{http://dx.doi.org/10.1007/JHEP11(2020)114}{{\em Journal of
  High Energy Physics} {\bfseries 2020} no.~11, (Nov., 2020) 114},
  \href{http://arxiv.org/abs/2002.00966}{{\ttfamily arXiv:2002.00966
  [hep-th]}}.

\bibitem{perturbedent1}
V.~Rosenhaus and M.~Smolkin, ``{Entanglement Entropy: A Perturbative
  Calculation},'' \href{http://dx.doi.org/10.1007/JHEP12(2014)179}{{\em JHEP}
  {\bfseries 12} (2014) 179}, \href{http://arxiv.org/abs/1403.3733}{{\ttfamily
  arXiv:1403.3733 [hep-th]}}.

\bibitem{intro:quench1}
P.~Calabrese and J.~Cardy, ``{Quantum Quenches in Extended Systems},''
  \href{http://dx.doi.org/10.1088/1742-5468/2007/06/P06008}{{\em J. Stat.
  Mech.} {\bfseries 0706} (2007) P06008},
  \href{http://arxiv.org/abs/0704.1880}{{\ttfamily arXiv:0704.1880
  [cond-mat.stat-mech]}}.

\bibitem{intro:quench2}
P.~Basu and S.~R. Das, ``{Quantum Quench across a Holographic Critical
  Point},'' \href{http://dx.doi.org/10.1007/JHEP01(2012)103}{{\em JHEP}
  {\bfseries 01} (2012) 103}, \href{http://arxiv.org/abs/1109.3909}{{\ttfamily
  arXiv:1109.3909 [hep-th]}}.

\bibitem{intro:quench3}
A.~Buchel, L.~Lehner, R.~C. Myers, and A.~van Niekerk, ``{Quantum quenches of
  holographic plasmas},'' \href{http://dx.doi.org/10.1007/JHEP05(2013)067}{{\em
  JHEP} {\bfseries 05} (2013) 067},
  \href{http://arxiv.org/abs/1302.2924}{{\ttfamily arXiv:1302.2924 [hep-th]}}.

\bibitem{intro:quench4}
S.~R. Das, D.~A. Galante, and R.~C. Myers, ``{Universal scaling in fast quantum
  quenches in conformal field theories},''
  \href{http://dx.doi.org/10.1103/PhysRevLett.112.171601}{{\em Phys. Rev.
  Lett.} {\bfseries 112} (2014) 171601},
  \href{http://arxiv.org/abs/1401.0560}{{\ttfamily arXiv:1401.0560 [hep-th]}}.

\bibitem{intro:quench5}
S.~R. Das, D.~A. Galante, and R.~C. Myers, ``{Universality in fast quantum
  quenches},'' \href{http://dx.doi.org/10.1007/JHEP02(2015)167}{{\em JHEP}
  {\bfseries 02} (2015) 167}, \href{http://arxiv.org/abs/1411.7710}{{\ttfamily
  arXiv:1411.7710 [hep-th]}}.

\bibitem{intro:quench6}
S.~R. Das, D.~A. Galante, and R.~C. Myers, ``{Smooth and fast versus
  instantaneous quenches in quantum field theory},''
  \href{http://dx.doi.org/10.1007/JHEP08(2015)073}{{\em JHEP} {\bfseries 08}
  (2015) 073}, \href{http://arxiv.org/abs/1505.05224}{{\ttfamily
  arXiv:1505.05224 [hep-th]}}.

\bibitem{intro:quench7}
S.~R. Das, D.~A. Galante, and R.~C. Myers, ``{Quantum Quenches in Free Field
  Theory: Universal Scaling at Any Rate},''
  \href{http://dx.doi.org/10.1007/JHEP05(2016)164}{{\em JHEP} {\bfseries 05}
  (2016) 164}, \href{http://arxiv.org/abs/1602.08547}{{\ttfamily
  arXiv:1602.08547 [hep-th]}}.

\bibitem{intro:quench8}
V.~{Alba} and P.~{Calabrese}, ``{Entanglement dynamics after quantum quenches
  in generic integrable systems},''
  \href{http://dx.doi.org/10.21468/SciPostPhys.4.3.017}{{\em SciPost Physics}
  {\bfseries 4} no.~3, (Mar., 2018) 017},
  \href{http://arxiv.org/abs/1712.07529}{{\ttfamily arXiv:1712.07529
  [cond-mat.stat-mech]}}.

\bibitem{Ghosh:2017nlk}
S.~Ghosh, K.~S. Gupta, and S.~C.~L. Srivastava, ``{Entanglement dynamics
  following a sudden quench: An exact solution},''
  \href{http://dx.doi.org/10.1209/0295-5075/120/50005}{{\em EPL} {\bfseries
  120} no.~5, (2017) 50005}, \href{http://arxiv.org/abs/1709.02202}{{\ttfamily
  arXiv:1709.02202 [quant-ph]}}.

\bibitem{Ghosh:2019yjh}
S.~Ghosh, K.~S. Gupta, and S.~C.~L. Srivastava, ``{Exact relaxation dynamics
  and quantum information scrambling in multiply quenched harmonic chains},''
  \href{http://dx.doi.org/10.1103/PhysRevE.100.012215}{{\em Phys. Rev. E}
  {\bfseries 100} no.~1, (2019) 012215},
  \href{http://arxiv.org/abs/1905.06743}{{\ttfamily arXiv:1905.06743
  [quant-ph]}}.

\bibitem{intro:ooe1}
A.~Polkovnikov, K.~Sengupta, A.~Silva, and M.~Vengalattore, ``Colloquium:
  Nonequilibrium dynamics of closed interacting quantum systems,''
  \href{http://dx.doi.org/10.1103/RevModPhys.83.863}{{\em Rev. Mod. Phys.}
  {\bfseries 83} (Aug, 2011) 863--883}.

\bibitem{intro:ooe2}
C.~Gogolin and J.~Eisert, ``Equilibration, thermalisation, and the emergence of
  statistical mechanics in closed quantum systems,''
  \href{http://dx.doi.org/10.1088/0034-4885/79/5/056001}{{\em Reports on
  Progress in Physics} {\bfseries 79} no.~5, (Apr, 2016) 056001}.

\bibitem{intro:ooe3}
P.~Calabrese, F.~H.~L. Essler, and G.~Mussardo, ``Introduction to `quantum
  integrability in out of equilibrium systems',''
  \href{http://dx.doi.org/10.1088/1742-5468/2016/06/064001}{{\em Journal of
  Statistical Mechanics: Theory and Experiment} {\bfseries 2016} no.~6, (Jun,
  2016) 064001}.

\bibitem{e1}
T.~Langen, T.~Gasenzer, and J.~Schmiedmayer, ``Prethermalization and universal
  dynamics in near-integrable quantum systems,''
  \href{http://dx.doi.org/10.1088/1742-5468/2016/06/064009}{{\em Journal of
  Statistical Mechanics: Theory and Experiment} {\bfseries 2016} no.~6, (Jun,
  2016) 064009}.

\bibitem{e2}
T.~Kinoshita, T.~Wenger, and D.~Weiss, ``A quantum newton's cradle,''
  \href{http://dx.doi.org/10.1038/nature04693}{{\em Nature} {\bfseries 440}
  (05, 2006) 900--3}.

\bibitem{e3}
S.~{Hofferberth}, I.~{Lesanovsky}, B.~{Fischer}, T.~{Schumm}, and
  J.~{Schmiedmayer}, ``{Non-equilibrium coherence dynamics in one-dimensional
  Bose gases},'' \href{http://dx.doi.org/10.1038/nature06149}{{\em \nat}
  {\bfseries 449} no.~7160, (Sept., 2007) 324--327},
  \href{http://arxiv.org/abs/0706.2259}{{\ttfamily arXiv:0706.2259
  [cond-mat.other]}}.

\bibitem{e4}
S.~{Trotzky}, Y.~A. {Chen}, A.~{Flesch}, I.~P. {McCulloch},
  U.~{Schollw{\"o}ck}, J.~{Eisert}, and I.~{Bloch}, ``{Probing the relaxation
  towards equilibrium in an isolated strongly correlated one-dimensional Bose
  gas},'' \href{http://dx.doi.org/10.1038/nphys2232}{{\em Nature Physics}
  {\bfseries 8} no.~4, (Apr., 2012) 325--330},
  \href{http://arxiv.org/abs/1101.2659}{{\ttfamily arXiv:1101.2659
  [cond-mat.quant-gas]}}.

\bibitem{e5}
M.~Gring, M.~Kuhnert, T.~Langen, T.~Kitagawa, B.~Rauer, M.~Schreitl, I.~Mazets,
  D.~A. Smith, E.~Demler, and J.~Schmiedmayer, ``Relaxation and
  prethermalization in an isolated quantum system,''
  \href{http://dx.doi.org/10.1126/science.1224953}{{\em Science} {\bfseries
  337} no.~6100, (2012) 1318--1322}.

\bibitem{e6}
M.~{Cheneau}, P.~{Barmettler}, D.~{Poletti}, M.~{Endres}, P.~{Schau{\ss}},
  T.~{Fukuhara}, C.~{Gross}, I.~{Bloch}, C.~{Kollath}, and S.~{Kuhr},
  ``{Light-cone-like spreading of correlations in a quantum many-body
  system},'' \href{http://dx.doi.org/10.1038/nature10748}{{\em \nat} {\bfseries
  481} no.~7382, (Jan., 2012) 484--487},
  \href{http://arxiv.org/abs/1111.0776}{{\ttfamily arXiv:1111.0776
  [cond-mat.quant-gas]}}.

\bibitem{e7}
F.~Meinert, M.~J. Mark, E.~Kirilov, K.~Lauber, P.~Weinmann, A.~J. Daley, and
  H.-C. N\"agerl, ``Quantum quench in an atomic one-dimensional ising chain,''
  \href{http://dx.doi.org/10.1103/PhysRevLett.111.053003}{{\em Phys. Rev.
  Lett.} {\bfseries 111} (Jul, 2013) 053003}.

\bibitem{e8}
T.~{Langen}, R.~{Geiger}, M.~{Kuhnert}, B.~{Rauer}, and J.~{Schmiedmayer},
  ``{Local emergence of thermal correlations in an isolated quantum many-body
  system},'' \href{http://dx.doi.org/10.1038/nphys2739}{{\em Nature Physics}
  {\bfseries 9} no.~10, (Oct., 2013) 640--643},
  \href{http://arxiv.org/abs/1305.3708}{{\ttfamily arXiv:1305.3708
  [cond-mat.quant-gas]}}.

\bibitem{e9}
T.~{Fukuhara}, P.~{Schau{\ss}}, M.~{Endres}, S.~{Hild}, M.~{Cheneau},
  I.~{Bloch}, and C.~{Gross}, ``{Microscopic observation of magnon bound states
  and their dynamics},'' \href{http://dx.doi.org/10.1038/nature12541}{{\em
  \nat} {\bfseries 502} no.~7469, (Oct., 2013) 76--79},
  \href{http://arxiv.org/abs/1305.6598}{{\ttfamily arXiv:1305.6598
  [cond-mat.quant-gas]}}.

\bibitem{e10}
T.~{Fukuhara}, A.~{Kantian}, M.~{Endres}, M.~{Cheneau}, P.~{Schau{\ss}},
  S.~{Hild}, D.~{Bellem}, U.~{Schollw{\"o}ck}, T.~{Giamarchi}, C.~{Gross},
  I.~{Bloch}, and S.~{Kuhr}, ``{Quantum dynamics of a mobile spin impurity},''
  \href{http://dx.doi.org/10.1038/nphys2561}{{\em Nature Physics} {\bfseries 9}
  no.~4, (Apr., 2013) 235--241},
  \href{http://arxiv.org/abs/1209.6468}{{\ttfamily arXiv:1209.6468
  [cond-mat.quant-gas]}}.

\bibitem{intro:quenchInt1}
S.~Sotiriadis and J.~Cardy, ``{Quantum quench in interacting field theory: A
  Self-consistent approximation},''
  \href{http://dx.doi.org/10.1103/PhysRevB.81.134305}{{\em Phys. Rev. B}
  {\bfseries 81} (2010) 134305},
  \href{http://arxiv.org/abs/1002.0167}{{\ttfamily arXiv:1002.0167
  [quant-ph]}}.

\bibitem{Coup11}
D.~Han, Y.~Kim, and M.~Noz, ``Illustrative example of feynman's rest of the
  universe,'' \href{http://dx.doi.org/10.1119/1.19192}{{\em American Journal of
  Physics - AMER J PHYS} {\bfseries 67} (01, 1999) }.

\bibitem{Coup12}
J.~S. Prauzner-Bechcicki, ``Two-mode squeezed vacuum state coupled to the
  common thermal reservoir,''
  \href{http://dx.doi.org/10.1088/0305-4470/37/15/l04}{{\em Journal of Physics
  A: Mathematical and General} {\bfseries 37} no.~15, (Mar, 2004) L173--L181}.

\bibitem{Coup13}
D.~Han, Y.~S. Kim, and M.~E. Noz, ``Linear canonical transformations of
  coherent and squeezed states in the wigner phase space. iii. two-mode
  states,'' \href{http://dx.doi.org/10.1103/PhysRevA.41.6233}{{\em Phys. Rev.
  A} {\bfseries 41} (Jun, 1990) 6233--6244}.

\bibitem{Coup14}
Y.~S. Kim, ``Observable gauge transformations in the parton picture,''
  \href{http://dx.doi.org/10.1103/PhysRevLett.63.348}{{\em Phys. Rev. Lett.}
  {\bfseries 63} (Jul, 1989) 348--351}.

\bibitem{Coup15}
F.~Iachello and S.~Oss, ``Model of n coupled anharmonic oscillators and
  applications to octahedral molecules,''
  \href{http://dx.doi.org/10.1103/PhysRevLett.66.2976}{{\em Phys. Rev. Lett.}
  {\bfseries 66} (Jun, 1991) 2976--2979}.

\bibitem{Coup21}
S.~Ikeda and F.~Fillaux, ``Incoherent elastic-neutron-scattering study of the
  vibrational dynamics and spin-related symmetry of protons in the
  ${\mathrm{khco}}_{3}$ crystal,''
  \href{http://dx.doi.org/10.1103/PhysRevB.59.4134}{{\em Phys. Rev. B}
  {\bfseries 59} (Feb, 1999) 4134--4145}.

\bibitem{COup22}
F.~Fillaux, ``Quantum entanglement and nonlocal proton transfer dynamics in
  dimers of formic acid and analogues,''
  \href{http://dx.doi.org/10.1016/j.cplett.2005.04.069}{{\em Chemical Physics
  Letters} {\bfseries 408} (06, 2005) 302--306}.

\bibitem{Coup23}
M.~Delor, S.~Archer, T.~Keane, A.~Meijer, I.~Sazanovich, G.~Greetham,
  M.~Towrie, and J.~Weinstein, ``Directing the path of light-induced electron
  transfer at a molecular fork using vibrational excitation,''
  \href{http://dx.doi.org/10.1038/nchem.2793}{{\em Nature Chemistry} {\bfseries
  9} (06, 2017) }.

\bibitem{Coup31}
E.~Romero, R.~Augulis, V.~Novoderezhkin, M.~Ferretti, J.~Thieme, D.~Zigmantas,
  and R.~van Grondelle, ``Quantum coherence in photosynthesis for efficient
  solar energy conversion,'' \href{http://dx.doi.org/10.1038/nphys3017}{{\em
  Nature Physics} {\bfseries 10} (07, 2014) }.

\bibitem{COup32}
F.~D. Fuller, J.~Pan, A.~Gelzinis, V.~Butkus, S.~S. Senlik, D.~E. Wilcox, C.~F.
  Yocum, L.~Valkunas, D.~Abramavicius, and J.~P. Ogilvie, ``Vibronic coherence
  in oxygenic photosynthesis,'' {\em Nature Chemistry} {\bfseries 6} no.~8,
  (Aug, 2014) 706--711.

\bibitem{Coup33}
A.~Halpin, P.~Johnson, R.~Tempelaar, R.~Murphy, J.~Knoester, T.~Jansen, and
  R.~Miller, ``Two-dimensional spectroscopy of a molecular dimer unveils the
  effects of vibronic coupling on exciton coherences,''
  \href{http://dx.doi.org/10.1038/nchem.1834}{{\em Nature chemistry} {\bfseries
  6} (03, 2014) 196--201}.

\bibitem{Coupe1}
D.~N. {Makarov}, ``{Coupled harmonic oscillators and their quantum
  entanglement},'' \href{http://dx.doi.org/10.1103/PhysRevE.97.042203}{{\em
  \pre} {\bfseries 97} no.~4, (Apr., 2018) 042203},
  \href{http://arxiv.org/abs/1710.01158}{{\ttfamily arXiv:1710.01158
  [quant-ph]}}.

\bibitem{Coupe2}
J.-Y. Kao and C.-H. Chou, ``Quantum entanglement in coupled harmonic oscillator
  systems: from micro to macro,''
  \href{http://dx.doi.org/10.1088/1367-2630/18/7/073001}{{\em New Journal of
  Physics} {\bfseries 18} no.~7, (Jul, 2016) 073001}.

\bibitem{Coupe3}
A.~{Jellal}, F.~{Madouri}, and A.~{Merdaci}, ``{Entanglement in coupled
  harmonic oscillators studied using a unitary transformation},''
  \href{http://dx.doi.org/10.1088/1742-5468/2011/09/P09015}{{\em Journal of
  Statistical Mechanics: Theory and Experiment} {\bfseries 2011} no.~9, (Sept.,
  2011) 09015}, \href{http://arxiv.org/abs/1106.3894}{{\ttfamily
  arXiv:1106.3894 [quant-ph]}}.

\bibitem{Coupe4}
M.~S. Abdalla, M.~Abdel-Aty, and A.-S.~F. Obada, ``Degree of entanglement for
  anisotropic coupled oscillators interacting with a single atom,''
  \href{http://dx.doi.org/10.1088/1464-4266/4/6/305}{{\em Journal of Optics B:
  Quantum and Semiclassical Optics} {\bfseries 4} no.~6, (Oct, 2002) 396--401}.

\bibitem{Coupe5}
M.~B. Plenio, J.~Hartley, and J.~Eisert, ``Dynamics and manipulation of
  entanglement in coupled harmonic systems with many degrees of freedom,''
  \href{http://dx.doi.org/10.1088/1367-2630/6/1/036}{{\em New Journal of
  Physics} {\bfseries 6} (Mar, 2004) 36--36}.

\bibitem{fieldH}
R.~A. {Jefferson} and R.~C. {Myers}, ``{Circuit complexity in quantum field
  theory},'' \href{http://dx.doi.org/10.1007/JHEP10(2017)107}{{\em Journal of
  High Energy Physics} {\bfseries 2017} no.~10, (Oct., 2017) 107},
  \href{http://arxiv.org/abs/1707.08570}{{\ttfamily arXiv:1707.08570
  [hep-th]}}.

\bibitem{Caputa:2017ixa}
P.~Caputa, S.~R. Das, M.~Nozaki, and A.~Tomiya, ``{Quantum Quench and Scaling
  of Entanglement Entropy},''
  \href{http://dx.doi.org/10.1016/j.physletb.2017.06.017}{{\em Phys. Lett. B}
  {\bfseries 772} (2017) 53--57},
  \href{http://arxiv.org/abs/1702.04359}{{\ttfamily arXiv:1702.04359
  [hep-th]}}.

\bibitem{PhysRevLett.122.081601}
H.~A. Camargo, P.~Caputa, D.~Das, M.~P. Heller, and R.~Jefferson, ``Complexity
  as a novel probe of quantum quenches: Universal scalings and purifications,''
  \href{http://dx.doi.org/10.1103/PhysRevLett.122.081601}{{\em Phys. Rev.
  Lett.} {\bfseries 122} (Feb, 2019) 081601}.

\bibitem{1994PhRvA..50.1035Y}
K.~H. {Yeon}, H.~J. {Kim}, C.~I. {Um}, T.~F. {George}, and L.~N. {Pandey},
  ``{Wave function in the invariant representation and squeezed-state function
  of the time-dependent harmonic oscillator},''
  \href{http://dx.doi.org/10.1103/PhysRevA.50.1035}{{\em \pra} {\bfseries 50}
  no.~2, (Aug., 1994) 1035--1039}.

\bibitem{cc1}
K.~Adhikari, S.~Choudhury, S.~Kumar, S.~Mandal, N.~Pandey, A.~Roy, S.~Sarkar,
  P.~Sarker, and S.~S. Shariff, ``{Circuit Complexity in
  $\mathcal{Z}_{2}$${\cal EEFT}$},''
  \href{http://arxiv.org/abs/2109.09759}{{\ttfamily arXiv:2109.09759
  [hep-th]}}.

\bibitem{cc2}
K.~Adhikari, S.~Choudhury, and A.~Roy, ``{${\cal K}$rylov ${\cal C}$omplexity
  in ${\cal Q}$uantum ${\cal F}$ield ${\cal T}$heory},''
  \href{http://arxiv.org/abs/2204.02250}{{\ttfamily arXiv:2204.02250
  [hep-th]}}.

\bibitem{cc3}
K.~Adhikari, S.~Choudhury, H.~N. Pandya, and R.~Srivastava, ``{PGW Circuit
  Complexity},'' \href{http://arxiv.org/abs/2108.10334}{{\ttfamily
  arXiv:2108.10334 [gr-qc]}}.

\bibitem{cc4}
S.~Choudhury, A.~Mukherjee, N.~Pandey, and A.~Roy, ``{Causality Constraint on
  Circuit Complexity from ${\cal COSMOEFT}$},''
  \href{http://arxiv.org/abs/2111.11468}{{\ttfamily arXiv:2111.11468
  [hep-th]}}.

\bibitem{cc5}
K.~Adhikari and S.~Choudhury, ``{${\cal C}$osmological ${\cal K}$rylov ${\cal
  C}$omplexity},'' \href{http://arxiv.org/abs/2203.14330}{{\ttfamily
  arXiv:2203.14330 [hep-th]}}.

\bibitem{cc6}
P.~Bhargava, S.~Choudhury, S.~Chowdhury, A.~Mishara, S.~P. Selvam, S.~Panda,
  and G.~D. Pasquino, ``{Quantum aspects of chaos and complexity from bouncing
  cosmology: A study with two-mode single field squeezed state formalism},''
  \href{http://dx.doi.org/10.21468/SciPostPhysCore.4.4.026}{{\em SciPost Phys.
  Core} {\bfseries 4} (2021) 026},
  \href{http://arxiv.org/abs/2009.03893}{{\ttfamily arXiv:2009.03893
  [hep-th]}}.

\bibitem{cc7}
R.~A. {Jefferson} and R.~C. {Myers}, ``{Circuit complexity in quantum field
  theory},'' \href{http://dx.doi.org/10.1007/JHEP10(2017)107}{{\em Journal of
  High Energy Physics} {\bfseries 2017} no.~10, (Oct., 2017) 107},
  \href{http://arxiv.org/abs/1707.08570}{{\ttfamily arXiv:1707.08570
  [hep-th]}}.

\bibitem{cent1}
K.~Adhikari, S.~Choudhury, S.~Chowdhury, K.~Shirish, and A.~Swain, ``{Circuit
  complexity as a novel probe of quantum entanglement: A study with black hole
  gas in arbitrary dimensions},''
  \href{http://dx.doi.org/10.1103/PhysRevD.104.065002}{{\em Phys. Rev. D}
  {\bfseries 104} no.~6, (2021) 065002},
  \href{http://arxiv.org/abs/2104.13940}{{\ttfamily arXiv:2104.13940
  [hep-th]}}.

\bibitem{cent2}
S.~Choudhury, S.~Chowdhury, N.~Gupta, A.~Mishara, S.~P. Selvam, S.~Panda, G.~D.
  Pasquino, C.~Singha, and A.~Swain, ``{Circuit Complexity from Cosmological
  Islands},'' \href{http://dx.doi.org/10.3390/sym13071301}{{\em Symmetry}
  {\bfseries 13} no.~7, (2021) 1301},
  \href{http://arxiv.org/abs/2012.10234}{{\ttfamily arXiv:2012.10234
  [hep-th]}}.

\bibitem{cent3}
J.~{Eisert}, ``{Entangling Power and Quantum Circuit Complexity},''
  \href{http://dx.doi.org/10.1103/PhysRevLett.127.020501}{{\em \prl} {\bfseries
  127} no.~2, (July, 2021) 020501},
  \href{http://arxiv.org/abs/2104.03332}{{\ttfamily arXiv:2104.03332
  [quant-ph]}}.

\bibitem{cent4}
S.~D. {Mathur}, ``{Three puzzles in cosmology},''
  \href{http://dx.doi.org/10.1142/S021827182030013X}{{\em International Journal
  of Modern Physics D} {\bfseries 29} no.~14, (Jan., 2020) 2030013},
  \href{http://arxiv.org/abs/2009.09832}{{\ttfamily arXiv:2009.09832
  [hep-th]}}.

\bibitem{cacv1}
D.~Stanford and L.~Susskind, ``Complexity and shock wave geometries,''
  \href{http://dx.doi.org/10.1103/PhysRevD.90.126007}{{\em Phys. Rev. D}
  {\bfseries 90} (Dec, 2014) 126007}.

\bibitem{cacv2}
L.~{Susskind}, ``{Addendum to Computational Complexity and Black Hole
  Horizons},'' {\em arXiv e-prints} (Mar., 2014) arXiv:1403.5695,
  \href{http://arxiv.org/abs/1403.5695}{{\ttfamily arXiv:1403.5695 [hep-th]}}.

\bibitem{cacv3}
D.~A. {Roberts}, D.~{Stanford}, and L.~{Susskind}, ``{Localized shocks},''
  \href{http://dx.doi.org/10.1007/JHEP03(2015)051}{{\em Journal of High Energy
  Physics} {\bfseries 2015} (Mar., 2015) 51},
  \href{http://arxiv.org/abs/1409.8180}{{\ttfamily arXiv:1409.8180 [hep-th]}}.

\bibitem{cacv4}
L.~{Susskind} and Y.~{Zhao}, ``{Switchbacks and the Bridge to Nowhere},'' {\em
  arXiv e-prints} (Aug., 2014) arXiv:1408.2823,
  \href{http://arxiv.org/abs/1408.2823}{{\ttfamily arXiv:1408.2823 [hep-th]}}.

\bibitem{2016arXiv160308747M}
S.~{Mukherjee}, A.~{Ghose Choudhury}, and P.~{Guha}, ``{Generalized damped
  Milne-Pinney equation and Chiellini method},'' {\em arXiv e-prints} (Mar.,
  2016) arXiv:1603.08747, \href{http://arxiv.org/abs/1603.08747}{{\ttfamily
  arXiv:1603.08747 [nlin.SI]}}.

\end{thebibliography}\endgroup
\bibliographystyle{utphys}

\end{document}